\documentclass[12pt]{iopart}
\usepackage{enumitem}
\usepackage{color}
\usepackage{colortbl}
\makeatletter
\edef\orig@output{\the\output}
\makeatother
\usepackage[utf8]{inputenc} 
\usepackage{hyperref}
\hypersetup{
    breaklinks=true,
    colorlinks=true,
    linkcolor=blue,
    citecolor=blue,      
    urlcolor=blue,
}

\usepackage[T1]{fontenc}
\usepackage{graphicx}
\usepackage[dvipsnames]{xcolor}
\usepackage{footnote}
\usepackage{bm}
\usepackage{cancel}
\usepackage{mathrsfs}
\usepackage{etoolbox}
\apptocmd{\sloppy}{\hbadness 10000\relax}{}{}
\usepackage{soul}
\usepackage[update,prepend]{epstopdf}
\usepackage{pifont}
\usepackage{cite}
\newcommand{\RN}[1]{%
  \textup{\uppercase\expandafter{\romannumeral#1}}%
}

\makesavenoteenv{tabular} 
\makesavenoteenv{table} 
\usepackage{multirow}
\renewcommand{\thesection}{\Roman{section}} 
\expandafter\let\csname equation*\endcsname\relax
\expandafter\let\csname endequation*\endcsname\relax
\usepackage{amsmath}
\usepackage{tikz,xcolor,hyperref}

\definecolor{lime}{HTML}{A6CE39}
\DeclareRobustCommand{\orcidicon}{%
	\begin{tikzpicture}
	\draw[lime, fill=lime] (0,0) 
	circle [radius=0.16] 
	node[white] {{\fontfamily{qag}\selectfont \tiny ID}};
	\draw[white, fill=white] (-0.0625,0.095) 
	circle [radius=0.007];
	\end{tikzpicture}
	\hspace{-2mm}
}

\foreach \x in {A, ..., Z}{%
	\expandafter\xdef\csname orcid\x\endcsname{\noexpand\href{https://orcid.org/\csname orcidauthor\x\endcsname}{\noexpand\orcidicon}}
}

\begin{document} 
\title[\textit{Csizmadia \textit{et al.} \NJP \textbf{23} 123012 (2021).}]{Detailed study of quantum path interferences in high harmonic generation driven by chirped laser pulses}

\author{Tamás Csizmadia{$^{1,}$\orcidJ{}\ding{41}}, Lénárd~Gulyás~Oldal{$^1$\orcidI{}}, Peng~Ye{$^1$\orcidH{}}, Szilárd~Majorosi{$^1$}, Paraskevas~Tzallas{$^{1,2}$\orcidF{}}, Giuseppe~Sansone{$^{1,3}$\orcidE{}}, Valer~Tosa{$^4$\orcidD{}}, Katalin~Varjú{$^{1,5}$\orcidC{}}, Balázs~Major{$^{1,5}$}\orcidB{}, Subhendu~Kahaly{$^{1,6,}$\orcidA{}\ding{41}}}
\address{$^1$ELI-ALPS, ELI-HU Non-Profit Ltd., Wolfgang Sandner utca 3., H-6728 Szeged, Hungary}
\address{$^2$Foundation for Research and Technology-Hellas, Institute of Electronic Structure \& Laser, GR-70013 Heraklion (Crete), Greece}
\address{$^3$Physikalisches Institut, Albert-Ludwigs-Universität, Stefan Meier Strasse 19, 79104 Freiburg, Germany}
\address{$^4$National Institute for R\&D of Isotopic and Molecular Technologies, RO, 400293, Cluj-Napoca, Romania}
\address{$^5$Department of Optics and Quantum Electronics, University of Szeged, Dóm tér 9, H-6728 Szeged, Hungary}
\address{$^6$Institute of Physics, University of Szeged, Dóm tér 9, H-6720 Szeged, Hungary}
\eads{\mailto{\ding{41}tamas.csizmadia@eli-alps.hu} and \mailto{\ding{41}subhendu.kahaly@eli-alps.hu}}

\begin{abstract}
We investigate the electron quantum path interference effects during high harmonic generation in atomic gas medium driven by ultrashort chirped laser pulses. To achieve that, we identify and vary the different experimentally relevant control parameters of such a driving laser pulse influencing the high harmonic spectra. Specifically, the impact of the pulse duration (from the few-cycle to the multi-cycle domain), peak intensity and instantaneous frequency is studied in a self-consistent manner. Simulations involving macroscopic propagation effects are also considered. The study aims to reveal the microscopic background behind a variety of interference patterns capturing important information both about the fundamental laser field and the generation process itself. The results provide guidance towards experiments with chirp control as a tool to unravel, explain and utilize the rich and complex interplay between quantum path interferences including the tuning of the periodicity of the intensity dependent oscillation of the harmonic signal, and the curvature of spectrally resolved Maker fringes.
\end{abstract}
 
\raggedbottom
\maketitle
\section{Introduction}
Coherent light sources in the extreme ultraviolet (XUV) and soft x-ray regimes is considered as one of the most useful resources to observe and manipulate various physical, chemical, and biological systems at their natural spatial (nanometric) and temporal (attosecond) scales \cite{Krausz2009,Nayak2019,Maroju2020}. Due to their relative compactness and low implementation cost, sources based on high harmonic generation in gases (GHHG) \cite{McPherson1987,Ferray1988} are the front-runners in generating coherent electromagnetic radiation in the short wavelength domain \cite{Kuhn2017,Charalambidis2017,Major2018,Ye2020}. Being a nonlinear optical process, GHHG can provide a broad spectrum and therefore ultrashort optical pulses in the form of attosecond pulse trains (APT) \cite{Paul2001} or even single attosecond pulses (SAP) \cite{Carrera2006,Goulielmakis2008}. Characterization and control of the intensity, temporal spacing, and duration of the individual atto-pulses constituting an APT are of utmost importance in attosecond physics and its applications \cite{Lpine2013,Kim2014,Orfanos2020}. These properties can be accessed through the investigation and shaping of the XUV spectral (or equivalently temporal) phase behaviour by the simultaneous tuning of specific parameters of the intense ultrashort laser pulse that drives the high harmonic generation process \cite{Chatziathanasiou2017}. In particular, the effect of the fundamental pulse’s chirp on GHHG process has been the subject of studies for several years \cite{Zhang1998,Lee2001}, and is still actively pursued \cite{Astiaso2016,Peng2018}. The harmonic spectral phase is inherently determined by the microscopic generation process theoretically described by the response of an individual atomic dipole to the strong alternating electric field. A~simple physical picture of this interaction is provided by the classical model that is built up of three distinguishable steps \cite{Corkum1993}: ($\emph{1}$) the emission of a single electron into the continuum (\textit{ionization}); ($\emph{2}$) \textit{acceleration} in the continuum over a classical trajectory drawn by the rapidly changing driving electric force and ($\emph{3}$) \textit{recombination}, in which the total (kinetic and potential) energy of the electron is converted into harmonic radiation. Although the classical model gives an intuitive insight into the dynamics of the electron under the influence of the intense laser field and provides good agreement for many experimentally observed features of the harmonic spectrum \cite{Krause1992}, it does not consider phenomena with quantum mechanical origin, like phase accumulation due to tunnelling, quantum diffusion of the electron wave packet during propagation, or quantum interference effects. A more realistic semi-classical model was developed by Lewenstein et al. \cite{Lewenstein1994} incorporating the Strong Field Approximation (SFA), where the influence of the atomic potential during the propagation of the electron wave packet in the continuum is neglected. The Lewenstein theory obtains the time-dependent dipole moment as
\begin{equation} \label{Lew}
\mathbf{D}(t)=i\int\limits^{t}_{-\infty}dt'\int\limits^{\infty}_{-\infty} d\mathbf{p}\mathbf{E}(t')\mathbf{d}(\mathbf{p}+\mathbf{A}(t')) \mathbf{d}^*(\mathbf{p}+\mathbf{A}(t))\mathrm{exp}\{iS(\mathbf{p},t,t')\}+c.c. ,
\end{equation}
In (\ref{Lew}) (expressed in atomic units, assuming a negative electron charge), $\mathbf{A}(t)$ describes the magnetic vector potential interacting with the target atom;  $\mathbf{E}(t)$ is the associated electric field ($\mathbf{E}(t)=-\partial{\mathbf{A}(t)}/\partial{t}$); $\mathbf{p}$ is the canonical momentum; $\mathbf{d(\mathbf{p})}$ is the dipole matrix element for bound-free transitions; and $S(\mathbf{p},t,t')$ is the quasiclassical action, given by the following expression:
\begin{equation} \label{Action}
S(\mathbf{p},t,t')=\int\limits^{t}_{t'}dt''\left(\frac{[\mathbf{p}+A(t'')]^2}{2}+I_\mathrm{p}\right) ,
\end{equation}
where $I_\mathrm{p}$ is the ionization energy. The Fourier transform of (\ref{Lew}), $\mathbf{D}(\omega_q)=\mathscr{F}\{\mathbf{D}(t)\}$ represents the emitted harmonic spectrum, where the emission rate associated to the harmonic frequency $\omega_q=q\omega_0$ ($\omega_0$ is the central angular frequency of the driving electric field) is proportional to $S(\omega_q)=\omega_q^3|D(\omega_q)|^2$. The Saddle Point Approximation (or Stationary Phase Approximation, SPA) allows us to convert the continuous integral within (\ref{Lew}) to a coherent superposition of quantum paths, which enables the separate investigation of the behaviour and contribution of different electron trajectories within a single emission process \cite{Nayak2019,Sansone2006}. Here, the classical electron paths are generalized to complex-valued functions (quantum trajectories), which consist of phase terms equal to the classical actions along the path and suitably assigned amplitudes. When the contributions of various trajectories become comparable, their relative phase difference introduces interference effects (quantum path interference, QPI) that play a cardinal role in the formation of the high harmonic spectrum, and can thereby serve as a special tool for probing the atomic dipole and allowing access to the electronic structure and dynamics of atoms or molecules \cite{Sansone2004,Varju2009,Yost2009}. Commonly, high harmonic generation setups can be considered as atomic scale, ultrafast interferometers, where the phase difference between electron trajectories can be altered with the spatial and temporal driving intensity $I(\mathbf{r},t)$, the emitted frequency $\omega_q$, the angle of emission $\theta$, as well as by various phase matching conditions in the generating medium \cite{Pedatzur2015,Azoury2018}. This physical complexity gives rise to a large variety of experimentally observable interference phenomena.\par
A simple way to directly perceive QPI is, for example, through the oscillation of the harmonic emission yield either as a result of changing the intensity \cite{Zair2008} or the wavelength of the driving electric field \cite{Schiessl2007}. Over the past decade, QPI has been extensively studied both in theory and throughout experiments using spectral and spatial filtering techniques \cite{Kanai2005, Cormier2009, Seres2015}, which has led to powerful applications in attosecond metrology targeting the precise in situ microscopic level control of the XUV emission process in atoms \cite{Kolliopoulos2014,Chatziathanasiou2019} or in aligned molecules \cite{Chatziathanasiou2019_2}. Liu et al. theoretically investigated QPI during HHG in argon driven by chirped laser pulses and demonstrated the multi-quantum path interference (MQPI) effect both in a single atom and in macroscopic medium \cite{Liu2009}. Recently, detailed interference structures were experimentally observed using chirped, multi-cycle laser pulses: Carlstr\"{o}m et al. discovered off-axis interference from contributions of long trajectories only, and on-axis interference with the participation of both short and long trajectory groups \cite{Carlstrom2016}. Although these studies revealed, both theoretically and experimentally, the dependence of QPI on the chirp of the fundamental laser field, the microscopic details of the underlying electron dynamics influenced by the different chirp-connected features of the driving pulse have not yet been investigated with a comprehensive, self-consistent methodology. Moreover, these researches on QPI driven by chirped pulses so far was conducted on the long pulse ($>30$~fs) regime, although multiple studies have shown that the trajectory behaviour is significantly altered due to the cycle-to-cycle variation of the fundamental laser pulse \cite{Sansone2004,Haessler2014}. This implies that QPI interpretation needs to be handled with care for ultrashort few-cycle pulses, which are now widely available to the community both from table-top laboratory sources, as well as in large-scale facilities like ELI-ALPS.\par
The purpose of the present paper is to reveal the role of different microscopic mechanisms that result in quantum path interferences during HHG in noble gas atoms exposed to intense, chirped laser pulses. Specifically, we describe the effect of chirp induced, sub-cycle and multi-cycle variations of the driving light pulse on the interference dynamics. The paper is organized as follows: In Section \ref{Model} we define the simple formula of a chirped light pulse with the set of parameter ranges used in our analysis, and depict the details of the model used in the calculations. Section \ref{Results_TL}--\ref{Results_combined} presents the results obtained through extensive numerical simulations corresponding to pulses with various peak intensities, durations and instantaneous frequency sweeps. In particular, Section \ref{Results_TL} deals with the effect of the the pulse duration and high harmonic order on the intensity dependent QPI oscillation utilizing transform-limited pulses as a base scenario. Section \ref{Results_instfreq} discusses the additional effect of the instantaneous frequency change on the QPI pattern, while Section \ref{Results_combined} deals with a realistic chirp model with all aforementioned effects incorporated. Relevant macroscopic calculations are performed in Section \ref{macro}. In Section \ref{Discussion}, we discuss the observed alterations of the interference patterns as a function of the applied laser pulse parameters. Finally, Section \ref{Conclusions} reports our conclusions.

\section{Description of the chirp model}\label{Model}
A diversity of chirp models have been utilized in literature \cite{Praxmeyer2005,Diels2006,Nakajima2007,NakajimaCormier2007,Peng2009,Mackenroth2016_PRL} to emphasize and express different physical aspects through the calculations. An~appropriate description of the chirped light field that is experimentally accessible and relevant is important for correlating the computational results with experimental observations. In our case we define the electric field in the spectral domain ($\mathbf{E}(\omega)$) so that the spectral intensity ($I(\omega)\propto|\mathbf{E}(\omega)|^2$, including both the peak spectral intensity and the bandwidth) is assumed to be unaffected by the chirp, a realistic scenario usually encountered in experiments, where the chirp is introduced by stretching an ultrashort pulse of a fixed energy:
\begin{equation} \label{E_omega}
\mathbf{E}(\omega)=\sqrt2\frac{\mathbf{E}_0c_B\pi}{\Delta\omega}\mathrm{exp}{\left(-\left(\frac{c_B\pi}{\Delta\omega}\right)^2(\omega-\omega_0)^2\right)}\mathrm{exp}(-i\phi(\omega)) ,
\end{equation}
where $\mathbf{E}_0$, and $\Delta\omega$ are the electric field amplitude and the full width at half maximum (FWHM) of the intensity ($I(\omega)\propto|\mathbf{E}(\omega)|^2$) envelope, respectively, and $c_B=\mathrm{2ln(2)/\pi}\approx0.4413$ is the time-bandwidth product for a pulse with a Gaussian envelope.\\
For ultrashort, chirped light pulses, the spectral phase term $\phi(\omega)$ in equation (\ref{E_omega}) is generally presumed to be a smooth function which can be expanded to a Taylor series around the $\omega_0$ central angular frequency as:
\begin{equation} \label{Spectral_phase}
 \phi(\omega)=\phi_0+\phi_1(\omega-\omega_0)+\frac{\phi_2}{2}(\omega-\omega_0)^2+\mathcal{O}(3).
\end{equation}
This can be approximated with the following formula:
\begin{equation}
    \phi(\omega)\approx-\phi_{\mathrm{CEO}}-\frac{1}{2}\mathrm{\arctan{(\xi)}}+\xi\left(\frac{c_B\pi}{\Delta\omega}\right)^2(\omega-\omega_0)^2+\mathcal{O}(3).
\end{equation}
Thereby, for consecutive Taylor coefficients one obtains:
\begin{equation}
    \phi_0=-\phi_{\mathrm{CEO}}-\frac{1}{2}\mathrm{\arctan{(\xi)},~}\phi_1=0\mathrm{,~and~} \phi_2=2\xi\left(\frac{c_B\pi}{\Delta\omega}\right)^2,
\end{equation}
where $\phi_2$ is the common physical quantity known as Group Delay Dispersion (GDD).
The Fourier transform of equation (\ref{E_omega}) (with $\mathcal{O}(3)$ phase terms neglected) becomes
\begin{equation} \label{E_t}
\mathbf{E}(t)=\frac{{\mathbf{E}_0}}{\sqrt[4]{(1+\xi^2)}}\mathrm{exp}{\left(-\frac{t^2}{\tau(\xi)^2}\right)}\mathrm{cos}(\xi\frac{t^{2}}{\tau(\xi)^2}+\omega_0t+\phi_{\mathrm{CEO}}),
\end{equation}
where $\tau (\xi)=\sqrt{1+\xi^2}\tau_0$ and $\tau_0=\frac{\tau_\mathrm{W,0}}{\sqrt{c_B\mathrm{\pi}}}$ are the chirped and the transform-limited pulse widths, respectively, and $\tau_\mathrm{W,0}$ is the full width at half maximum (FWHM) of the intensity ($I(t)\propto|\mathbf{E}(t)|^2$) envelope for $\xi=0$. The calculations were carried out with fields having $\omega_0=1.829$~rad/fs central angular frequency (corresponding to a 1030~nm central wavelength) and $\phi_{\mathrm{CEO}}=0$ carrier envelope offset. $\mathbf{E}(t)$~was considered to be linearly polarized. Equation (\ref{E_t}) explicitly shows that three major characteristics\textemdash the peak amplitude, the envelope, and the phase\textemdash of the transform-limited ($\xi=0$) driving field are altered by the presence of the chirp, all of which have significant and distinct impacts on QPI in HHG.
In order to benchmark, compare and differentiate between the effects on the HHG process caused by these distinguishable (but interconnected) chirp induced modifications of the driving pulse, we define three different scenarios, which could also be implemented in real-life experiments. Without any restriction, in general one can rewrite the electric field as:
\begin{equation} \label{E_t_distinguished}
\mathbf{E}(t)=\frac{\mathbf{E}_0}{\sqrt[4]{(1+\xi_{\RN{1}}^2)}}\mathrm{exp}{\left(-\frac{t^2}{(1+\xi_{\RN{2}}^2)\tau_0^2}\right)}
\mathrm{cos}\left(\xi_{\RN{3}}\frac{t^{2}}{(1+\xi_{\RN{3}}^2)\tau_0^2}+\omega_0t+\phi_{CEO}\right) ,
\end{equation}
where $\xi_\RN{1}$, $\xi_\RN{2}$ and $\xi_\RN{3}$ are three individual control parameters. It is evident from (\ref{E_t_distinguished}) that the control parameters can be appropriately tuned to probe different aspects of the QPI process:
\begin{enumerate}[label=\Roman*.]
\item The peak intensity changes as\footnote{Due to computational reasons, the vector potential $A(t)=-\int_{-\infty}^{t} E(t')dt'$ was approximated with the closed-form expression $A(t)=-\frac{E_0}{\omega_0}\frac{1}{\sqrt[4]{(1+\xi^2)}}\mathrm{exp}{\left(-\frac{t^2}{\tau(\xi)^2}\right)}\mathrm{sin}\left(\xi\frac{t^{2}}{\tau(\xi)^2}+\omega_0t\right)$ for calculating (\ref{Lew}). In this case, both the peak intensity (\ref{Int}) and pulse duration (\ref{Length}) slightly depend on $\xi_\RN{3}$. However, equations (\ref{Int}) and (\ref{Length}) generally provide a very good approximation of the peak intensity and pulse duration, respectively, with a relative error of $<5\times10^{-3}$ corresponding to the parameter range used in this work.}
\begin{equation}\label{Int}
I=I_0/\sqrt{1+\xi_{\RN{1}}^2} .
\end{equation}
\item The pulse width is modified as
\begin{equation}\label{Length}
\tau_\mathrm{W}=\sqrt{1+\xi_{\RN{2}}^2}\tau_\mathrm{W,0} .
\end{equation}
\item The instantaneous frequency within the pulse spans towards higher ($\xi_{\RN{3}}>0$) or lower ($\xi_{\RN{3}}<0$) frequencies within a defined frequency interval that depends on the chirp parameter. The difference between the instantaneous frequencies at the beginning and at the end of the temporal electric field can be expressed as:
\begin{equation} \label{Instfreq}
\delta\omega_{\mathrm{inst}}=\omega(t_\mathrm{max})-\omega(t_\mathrm{min})\propto\frac{\xi_{\RN{3}}}{\sqrt{1+\xi_{\RN{3}}^2}}\Delta\omega ,
\end{equation}
where $\Delta\omega=c_\mathrm{B}2\mathrm{\pi}/\tau_0$ is the FWHM spectral bandwidth, and $t_\mathrm{max}-t_\mathrm{min}=\sqrt{1+\xi_{\RN{3}}^2}\tau_\mathrm{0}$.
\begin{itemize}
\item If $\xi_{\RN{3}}=0$, $\omega(t_\mathrm{max})-\omega(t_\mathrm{min})=0$, the instantaneous frequency does not change in time, there is no chirp.
\item If $\left|\xi_{\RN{3}}\right|\ll1$, $\omega(t_\mathrm{max})-\omega(t_\mathrm{min})\propto\xi_\RN{3}$, the frequency shift linearly varies with the chirp parameter.
\item I $\left|\xi_{\RN{3}}\right|\gg1$, $\omega(t_\mathrm{max})-\omega(t_\mathrm{min})\propto\Delta\omega$. In this case, the maximum frequency variation covers the bandwidth that matches the known linear nature of chirp (new frequency components do not appear).
\end{itemize}
\end{enumerate}

Therefore, tuning $\xi_\RN{1}$ (with $\xi_\RN{2}=\xi_\RN{3}=0$) in (\ref{E_t_distinguished}) is equivalent to controlling the peak intensity of a transform-limited pulse with a given duration. Variation in $\xi_\RN{2}$ (with $\xi_\RN{1}=\xi_\RN{3}=0$) represents a change in transform-limited pulse duration with the same peak intensity, and finally $\xi$ ($=\xi_\RN{1}=\xi_\RN{2}=\xi_\RN{3}$) indicates the most general case with chirp $\xi$, where the peak intensity and pulse duration are self-consistently modified as mentioned before.
\begin{table}
\centering
\begin{tabular}{|c|c|c|c|c|} 
\hline
\rowcolor[gray]{.9}[0.80\tabcolsep]
\textbf{Process} & \textbf{Parameter} &  \textbf{From} &  \textbf{To} & \textbf{Unit} \\ [1ex] 
\hline\hline
& \textit{$\xi$} & 0 & $\pm$3 & normalized \\
\rowcolor[gray]{.95}[0.60\tabcolsep]
\textbf{(\RN{1})} & $I$ & $1.72\times10^{14}$ & $5.4\times10^{13}$ & Wcm$^{-2}$\\
\textbf{(\RN{2})} & $\tau_{\mathrm{W}}$ & 15.1 & 47.8 & fs \\
\rowcolor[gray]{.95}[0.60\tabcolsep]
\textbf{({\RN{3}})} & $\delta\omega_{\mathrm{inst}}$
\footnote{Calculated at $\xi_\RN{3}=3$ and $t_\mathrm{max}-t_\mathrm{min}=\tau_\mathrm{W}$ using (\ref{Instfreq}).} & 1.742 & 1.916 & rad/fs \\
\hline
\end{tabular}
\caption{The parameter range used in the calculations.}
\label{Parameters}
\end{table}
In order to investigate the effects of these contributing mechanisms separately, a specific parameter range has been selected (shown in Table \ref{Parameters}) corresponding to the same chirp range ($-3\leq\xi=\xi_{\RN{1}}=\xi_{\RN{2}}=\xi_{\RN{3}}\leq3$) in case of all three aforementioned actions. The total applied intensity range in Process~(\RN{1}) was chosen considering that it should support the generation of harmonics, while preferably keeping the maximum applied intensity ($I_0$) relatively low in order to avoid excessive ionization of the medium\footnote{Using the set of parameters in Table \ref{Parameters}, the maximum rate of ionization at focus (@$\xi$=0) would be 3.4\%, while @$\xi$=3 only 0.011\% of the atoms would be ionized.}\label{Footnote_ionization}, thereby preserving the validity of the single-atom simulations. The resulting intensity range defined the chirp parameter regime in Table \ref{Parameters} based on (\ref{Int}). Every computation involved reference atoms having a Gaussian ground state wave function ($\mathbf{d}(\mathbf{p})=i/(\pi\alpha_f)^{3/4}(\mathbf{p}/\alpha_f) \mathrm{exp}\left[-\mathbf{p}^2/(2\alpha_f)\right]$), where $\alpha_f=I_\mathrm{p}$ is the fitting parameter adjusted to the ionization potential of argon (15.76~eV)\cite{Lewenstein1994,Sansone2004,Nayak2019}). Recombinations after multiple visits of the parent ion by the accelerating electron have been neglected (therefore only the contributions of trajectories shorter than one optical cycle have been considered). We point out here that the atomic phase shift is neglected in the SFA and it can affect the high-order harmonic spectrum by introducing features that are dependent on the specific electronic structure of the target atom. The Cooper minimum (CM) in argon \cite{wahlstrom1993,Minemoto2008, Worner2009b} and other ionic \cite{Hassouneh2018} or molecular systems \cite{Scarborough2018} is one such instance. It is worth to note, however, that the location of CM in the high-order harmonic spectrum does not depend on the driving laser intensity and wavelength (see for example figure 1 in \cite{Worner2009b} and \cite{ Colosimo2008}) and the spectral phase variation of the harmonics near the CM is observed to be limited to a narrow region near the dip \cite{ Worner2009b,Schoun2014}. Thus, in general for systems exhibiting CM like behaviour or other intrinsic effects like giant resonance \cite{ Shiner2011b }, the experimental conditions can be adapted to either a lower intensity regime, so that the harmonics in the CM region does not reach the plateau of the spectrum, or one can spectrally filter and select the appropriate harmonic orders to record the QPI patterns, where CM effects do not appear.\\
To explain in more details the finer aspects of CM goes beyond the scope of this study. Nevertheless, QPI effects depend strongly on the phase accumulated by the electronic wave packet in the continuum, which is a strong function of the intensity and the temporal shape of the laser field in general supporting the validity of the SFA in studying QPI \cite{Zair2008,Zair2012,Schapper2010,Teichmann2012}. Hence, in a differential intensity resolved measurement, such as the one carried out in our work, one would expect that the QPI features would be retained.\par
Moreover, in order to ensure the validity of the simulations, various single-atom emission spectra calculated with the Lewenstein integral and the saddle point method \cite{Nayak2019} were compared. We also carried out numerical simulations solving the time-dependent Schr\"{o}dinger equation (TDSE) with a soft-core Coulomb potential in the single active electron (SAE) approximation \cite{Majorosi2018} in some cases to justify the correctness of the SFA calculations.

\section{QPI driven by transform-limited ultrashort pulses}
\label{Results_TL}
\subsection{Effect of the pulse duration and high harmonic order on the intensity dependent QPI oscillation}
\begin{figure*}[ht]
  \centering
  \includegraphics[width=0.85\linewidth]{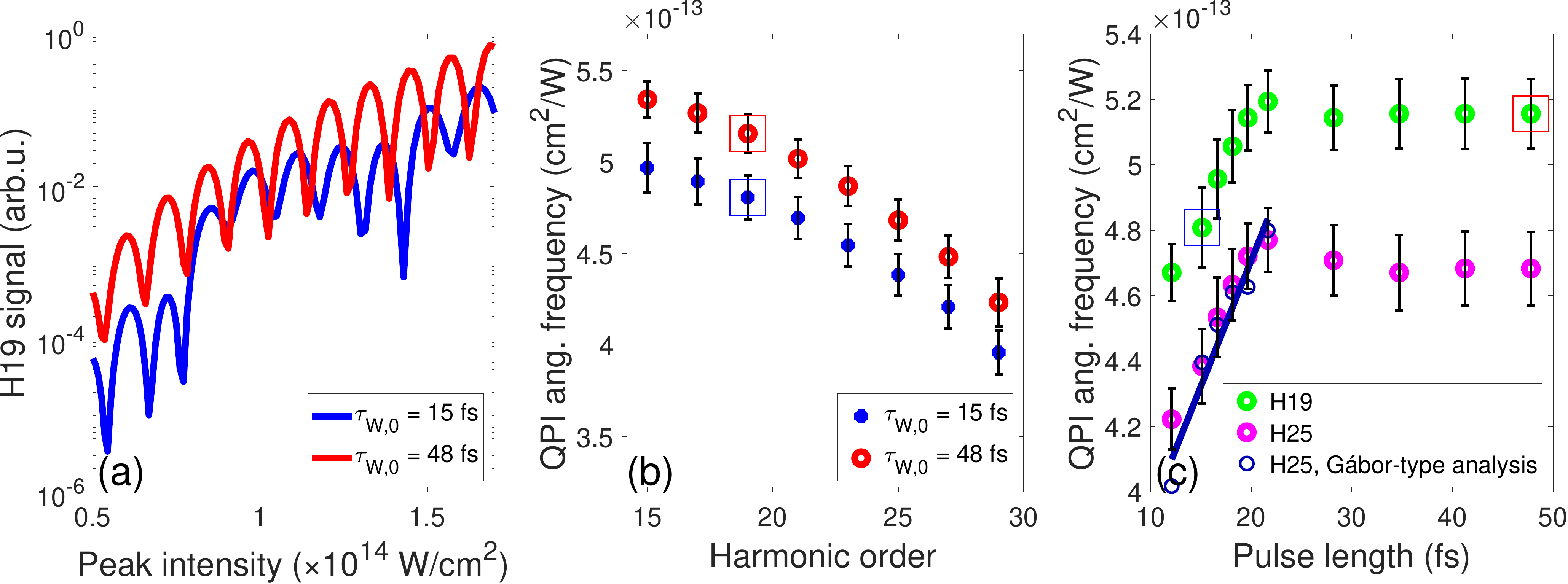}
  \caption{(a) Oscillation of the single-atom harmonic emission rate as a function of the peak laser intensity in case of two Fourier-limited pulses with different pulse lengths: $\tau_\mathrm{W,0}=15$~fs (blue) and $\tau_\mathrm{W,0}=48$~fs (red). (b) The dependence of the angular frequency of the QPI oscillation on the harmonic order. (c) Variation of the QPI angular frequency with the pulse length using Fourier-limited driving pulses in case of the 19\textsuperscript{th} (green) and 25\textsuperscript{th} (purple marks) harmonic orders. The blue (red) squares in figure \ref{Fig1}(b) and (c) indicate the same data extracted from the oscillating curve in case of the 15 (48)~fs long fields, both depicted in figure \ref{Fig1}(a).}
  \label{Fig1}
\end{figure*}

Following  the \textit{ceteris paribus} principle, first we investigate only Process (I) considering Fourier-limited driving fields (i.e. $\xi_{\RN{2}}$, and $\xi_{\RN{3}}=0$ in (\ref{E_t_distinguished})) with different pulse durations. Alteration in the peak intensity of the driving electric field alters QPI via the modification of the phases of particular trajectories associated to the same harmonic order. The phase $\phi^q_j$ associated to the quantum path $j$ contributing to the harmonic order $q$ is directly connected to the quasiclassical action (equation (\ref{Action})) and can be approximated by: $\phi^q_j\approx-U_\mathrm{p}\tau^q_j\approx-\alpha^q_jI$ (in atomic units), where $U_\mathrm{p}$ is the ponderomotive energy and $\alpha^q_j$, known as reciprocal intensity, is roughly proportional to the electron excursion time $\tau^q_j$ \cite{He2015}. Longer trajectories inherently result in stronger intensity dependence of the corresponding dipole phase. The interference between quantum paths with marginally different excursion times ("long" and "short" trajectories) results in an intensity dependent oscillation in the harmonic signal with the angular frequency of $\Delta\alpha^q=\alpha^q_l-\alpha^q_s$, correspondingly to their phase difference. This is demonstrated in figure \ref{Fig1}(a) for a plateau harmonic (19\textsuperscript{th}) in case of two unchirped pulses with different temporal extents: $\tau_\mathrm{W,0}=$15~fs (blue) and 48~fs (red). Clear oscillations are visible in case of both pulse lengths, although with different oscillation frequencies, $\Delta\alpha^\mathrm{H19}(\tau_\mathrm{W,0}$=15~fs)=4.81$\pm0.12\times10^{-13}$~cm$^2$W$^{-1}$ for the shorter (blue) pulse, and greater $\Delta\alpha^\mathrm{H19}(\tau_\mathrm{W,0}=48$~fs)=5.16$\pm0.11\times10^{-13}$~cm$^2$W$^{-1}$ for the longer (red) pulse. As the harmonic order $q$ increases, $\partial\phi^q_j/\partial I$ decreases (increases) for the short (long) trajectories in all half-cycles in the laser field, converging towards a common value in the cut-off region \cite{Varju2006}. Consequently, differences between $\phi^q_j$ of different $j$ trajectories become smaller, which results in a decrease of the QPI frequency as represented in figure \ref{Fig1}(b).\par
While preserving the oscillation introduced by Process (\RN{1}), we now incorporate the effect of Process (\RN{2}) in order to investigate the dependence of the QPI angular frequency on the transform-limited pulse duration. In this case, the observation in figure \ref{Fig1}(a) applies in figure \ref{Fig1}(b) as a general trend, i.e. higher QPI frequencies are produced with the longer driving field in case of all investigated harmonics. This effect is demonstrated in a broader pulse duration range, from $\tau_\mathrm{W,0}=12$~fs to $48$~fs in case of two plateau harmonics in figure \ref{Fig1}(c), where a clear change of the QPI periodicity can be observed for the pulse duration range between approximately 12 and 25~fs, followed by a relatively constant region for pulses longer than 25~fs. In order to gain insight into the underlying processes affecting the QPI frequency as a function of pulse duration and harmonic order, the contribution of different quantum trajectories (having $\alpha^q_j$ reciprocal intensities) to the formation of the dipole signal (oscillating with the frequency of $\Delta\alpha^q$) has to be investigated.\par

\subsection{Disentangling quantum path contributions without trajectory analysis}
\begin{figure*}[ht]
  \centering
  \includegraphics[width=0.9\linewidth]{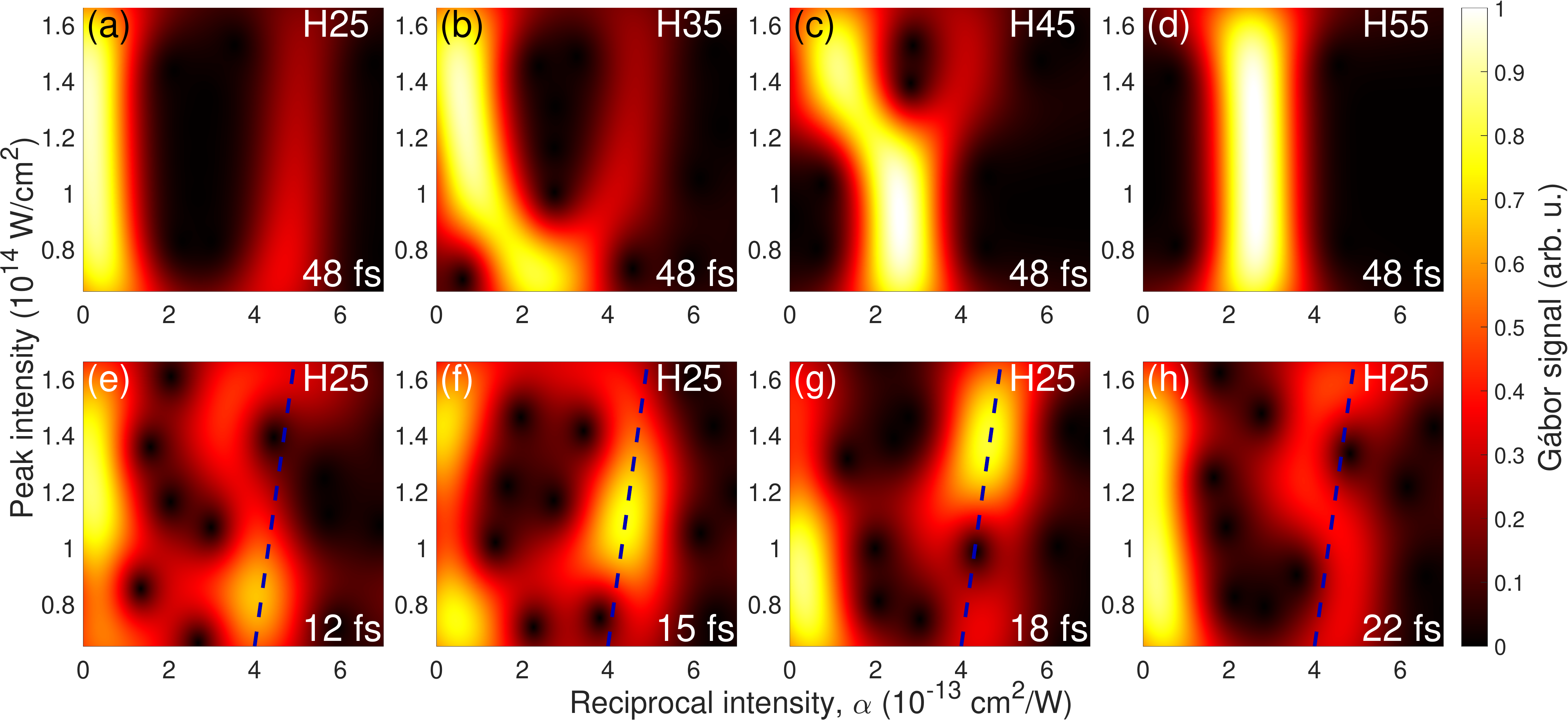}
  \caption{Pseudo-colour maps of the intensity dependent phase behaviour of different harmonic orders (25\textsuperscript{th} (a), 35\textsuperscript{th} (b), 45\textsuperscript{th} (c), and 55\textsuperscript{th} (d)), and in case of pulse lengths ($\tau_\mathrm{W,0}=12$~fs (e), 15~fs (f), 18~fs (g), and 22~fs (h)). A $\tau_\mathrm{W,0}=48$~fs long pulse was used to calculate subplots (a--d), while (e--h) were evaluated from the dipole phase corresponding to the 25\textsuperscript{th} harmonic.}
  \label{Fig2}
\end{figure*}

It is possible to disentangle the different quantum path contributions from the dipole moment for several laser peak intensities by using a wavelet-like sliding window Fourier transform on the intensity dependent dipole moment. This simple numerical method has already been utilized by several research groups \cite{Balcou1999,Gaarde1999,Auguste2009} and it can be viewed as a Gábor-type of analysis on an intensity dependent signal resulting in a 3D representation of the intensity ($I^\prime$) - reciprocal intensity ($\alpha^q$) plane. Specifically, the Fourier transform is executed on the apodized dipole term $d^q_{\phi}(I)$ consisting of the intensity dependent dipole phase at a given harmonic multiplied by a Gaussian window function:
\begin{equation}\label{Gabor_type}
    G(I^\prime,\alpha^q)=\mathscr{F}\{d^q_{\phi}(I)\}=\int\limits^{\infty}_{-\infty}e^{i\phi^q(I)}e^{-(a_w(I^\prime-I))^2}e^{-i\alpha^qI}dI ,
\end{equation}
where $\phi^q(I)=\mathrm{arg}(D(q,I))$, and $I^\prime$ is the central intensity for a given apodization window spanning from $I_1+1/a_w$ up to typically $I_2-1/a_w$, where $I_1$ and $I_2$ mark the lowest and highest $I$ intensities of the studied peak intensity range, respectively. The value $a_w$ corresponds to the length of the apodization window and is determined by a trade-off between the resolutions on the intensity and reciprocal intensity axes. In the calculations presented in figure \ref{Fig2}, $a_w=5\times10^{-14}$~cm$^2$W$^{-1}$ has been used. By choosing a sufficiently small $a_w$ value, (\ref{Gabor_type}) practically turns into a Fourier transform resulting in the QPI angular frequency values represented in figure \ref{Fig1}(b).\par
Figure \ref{Fig2}(a) shows that in case of the 25\textsuperscript{th} harmonic, which is located in the plateau region for the entire investigated intensity range, the intensity/reciprocal intensity distribution is dominated by two distinguishable trajectory classes with marginally different $\alpha^{H25}_j$ values, namely short paths (with $\tau_s^q<0.6167\times2\pi/\omega_0$ and $\alpha^{H25}_s<10^{-13}$~cm$^2$W$^{-1}$) and long paths (with $\tau_l^q>0.6167\times2\pi/\omega_0$ and $\alpha^{H25}_l\approx4$--$6\times10^{-13}$~cm$^2$W$^{-1}$). When the harmonic order is shifted towards higher values (figure \ref{Fig2}(b) and (c)), the $\alpha$ values of the two trajectory groups tend to move towards each other to join as a single class of cut-off trajectory (figure \ref{Fig2}(d)) resulting in a smaller $\Delta\alpha^q=\alpha^q_l-\alpha^q_s$ value. Thus this upward shift of the characteristic fork-like structure of the density plots explains the decrease in the QPI angular frequency observed in figure \ref{Fig1}(b).\par
Concerning the pulse length dependence shown in figure \ref{Fig1}(c), figure \ref{Fig2}(e) reveals that for $\tau_\mathrm{W,0}=12.1$~fs the contribution of the long trajectory class is strong around one particular intensity (at $I^\prime\approx0.85\times10^{14}$~Wcm$^{-2}$). At this intensity, the $\alpha^{\mathrm{H25}}_l$ value of the long trajectory (defined as the position of the signal maximum) can be extracted. When tuning the pulse length (figure \ref{Fig2} e--h)), the maximum is shifted along a linear curve (marked with a dashed dark blue line) until it exits from the studied peak intensity range at $\tau_\mathrm{W,0}=22$~fs (figure \ref{Fig2} h)), while the $\alpha^{H25}_s$ of the short trajectory remains relatively constant. The $\Delta\alpha^{\mathrm{H25}}$ values corresponding to different pulse durations are depicted in figure \ref{Fig1}(c) (with empty dark blue circles and a solid dark blue line as a linear fitting) showing good agreement with the previously calculated QPI angular frequencies.\par

\section{Effect of instantaneous frequency on QPI pattern}
\label{Results_instfreq}
\begin{figure*}[ht]
  \centering
  \includegraphics[width=0.95\linewidth]{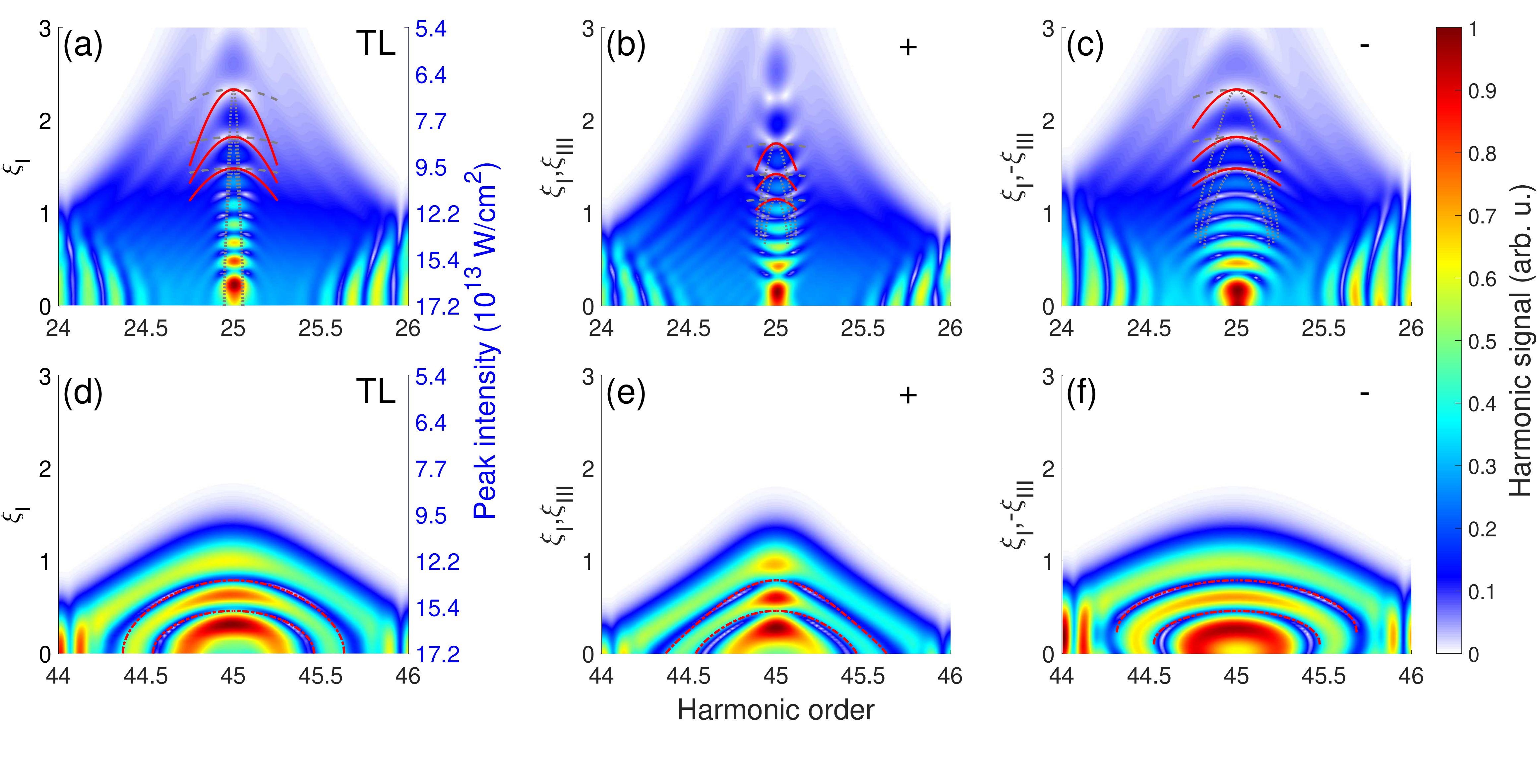}
  \caption{The harmonic spectra around the 25\textsuperscript{th} (a, b, c) and 45\textsuperscript{th} (d, e, f) harmonics as a function of the chirp parameter in case of (a, d) transform-limited ($\xi_{\RN{2},\RN{3}}=0$, $\xi_{\RN{1}}$ changes), (b, e) positively chirped ($\xi_{\RN{2}}=0$, $\xi_{\RN{1},\RN{3}}$ changes) and (c, f) negatively chirped ($\xi_{\RN{2}}=0$, $\xi_{\RN{1},\RN{3}}$ oppositely changes) driving pulses. 48~fs long driving pulses were utilized in all six cases. The same phase difference between short (grey dotted), long (grey dashed), and cut-off (red dot-and-dashed) trajectories were indicated with lines defined by (\ref{Maker_fit}) using the $\alpha^q$ value of the given trajectory. The red solid lines were calculated with a single intermediate $\alpha^q$ value ($\alpha^q_s<\alpha^q<\alpha^q_l$). The square root of the emission rate ($\sqrt{S_q(\omega_q)}$) was plotted (applying also to figures \ref{Fig4} and \ref{Fig7}) to enhance the visualization of the interference features in the evolving high harmonic spectra.}
  \label{Fig3}
\end{figure*}

In order to study the effect of the instantaneous frequency variation described in Process (\RN{3}), three intensity-scans ($0\leq\xi_{\RN{1}}\leq3$) were conducted with the use of Fourier-limited ($\xi_{\RN{2}}=0$), positively chirped ($0<\xi_{\RN{3}}\leq3$), and negatively chirped ($-3\leq\xi_{\RN{3}}<0$) fundamental fields shown in figures \ref{Fig3}(a) to (c) and d--e) for spectra around one lower plateau (25$^\mathrm{th}$) and one higher plateau harmonic (45$^\mathrm{th}$), respectively. The pulse length of the applied pulses was kept at $\tau_\mathrm{W,0}=48$~fs for all three scans in pursuit of excluding the effect of the pulse length change on the QPI pattern. The fact that $\tau_\mathrm{W,0}$ is fixed ($\xi_\RN{2}=0$ or constant) entails that the transform-limited duration is fixed, so the bandwidth is fixed. Figure \ref{Fig3} shows rich interference patterns in the spectral domain for all three cases. Since the Gábor-type transformation is not able to resolve the intra-pulse dynamics of the driving laser field needed to differentiate between positively and negatively chirped fundamental pulses, we analyze this case by studying the markedly different interference patterns visualized in figure \ref{Fig3}. Two main effects contribute to the apparent change of the observable textures. The instantaneous frequency of harmonic $q$ can be described as
\begin{equation}
\omega_q(t)=q\omega_0+\Delta\omega_{\mathrm{IR},q}(t)+\Delta\omega_\mathrm{dip}(t) ,
\label{Harm_freq}
\end{equation}
where $\Delta\omega_{\mathrm{IR},q}(t)$ and $\Delta\omega_\mathrm{dip}(t)$ are the contributions from the instantaneous frequency variation of the driving field and the harmonic chirp caused by the intensity dependence of the atomic dipole phase, respectively \cite{Gaarde1999,Nefedova2018}. At the rising (falling) edge of the pulse, where the intensity increases (decreases), $\Delta\omega_\mathrm{dip}(t)$ is always positive (negative), which leads to a blue (red) shift at this part of the pulse. These two shifts result in the broadening of the harmonics, provided that no ground state depletion takes place. The broadening is stronger for higher peak intensities (when $|\partial I/\partial t|$ is greater) at smaller $\xi_{\RN{1}}$ values, as shown in figure \ref{Fig3}(a). A positively chirped fundamental pulse contributes with a red frequency shift ($\Delta\omega_{\mathrm{IR},q}(t)<0$) during the rising edge and a blue shift ($\Delta\omega_{\mathrm{IR},q}(t)>0$) during the trailing edge that compensates the harmonic chirp resulting in sharp harmonics for high $\xi_{\RN{3}}$ values (see around $\xi_{\RN{3}}=3$ in figure \ref{Fig3}(b)). On the contrary, a negatively chirped driver further enhances the broadening effect introduced by the intensity dependent dipole phase, resulting in increasingly broadening harmonics with increasing $\xi_{\RN{3}}$ as observed in figure \ref{Fig3}(c). The frequency chirp of harmonics has been extensively studied both in theory \cite{Salieres1995,Kan1995,Kim2001} and in laboratory experiments \cite{Shin1999}.
In order to explain the shape of the interference pattern, the instantaneous frequency of harmonic $q$ was analytically calculated from (\ref{Harm_freq}), where $\Delta\omega_{\mathrm{IR},q}(t)=2qt\xi_\RN{3}/((1+\xi^2_\RN{3})\tau_0^2)$ and $\Delta\omega_\mathrm{dip}(t)=\alpha_j^q\partial I(t)/\partial t$. The use of (\ref{Action}) to express $\Delta\omega_\mathrm{dip}(t)$ results in the following equation:
\begin{equation}
\Delta\Omega=\omega_q(\xi_\RN{1},\xi_\RN{3})-q\omega_0=\pm \left(\frac{2q\xi_\RN{3}}{(1+\xi^2_\RN{3})\tau_0}-\alpha_j^q\frac{4I_m}{(1+\xi_\RN{2}^2)\tau_0}\right)\sqrt{\frac{1+\xi_\RN{2}^2}{2}\mathrm{ln}\frac{I_0}{\sqrt{1+\xi_\RN{1}^2}I_m}} ,
\label{Maker_fit}
\end{equation}
where $I_m$ marks certain intensities corresponding to given $t_n$ times, at which the same phase difference between trajectories are sustained for different pulse shapes depending on the applied chirp. In real experimental conditions, the $I_m$ term refers to intensities at which longitudinal phase matching conditions are optimized spatially and temporally during the simultaneous propagation of the IR and XUV fields, thereby macroscopic effects are also introduced. Due to the role of phase matching in the formation, these fringes are often referred to as Maker fringes \cite{Maker1962,Heyl2011,Catoire2016}, even though the shape of the fringe pattern can still be traced back to microscopic mechanisms \cite{Heyl2011,Catoire2016}. In our calculations, $I_m=I_0/\sqrt{1+\xi_{\RN{1},0}^2}$, where $\xi_{\RN{1},0}$ marks the starting points of the curves corresponding to $\Delta\Omega(\xi_\RN{1},\xi_\RN{3})=0$.
In figure \ref{Fig3}, the lines represent $\xi_{\RN{1}}$ and $\xi_{\RN{3}}$ as a function of $\Delta\Omega$ (assuming  $\xi_{\RN{2}}=0$ in all cases). Figures \ref{Fig3}(a), (b), and (c) show lines calculated for the short (grey dotted line), and long trajectories (grey dashed line) at three arbitrary $\xi_{\RN{1},0}$ values using the $\alpha^{\mathrm{H25}}_j$ values extracted from the Gábor-type analysis depicted in figure \ref{Fig2}. The spectral interference fringes visibly fall between the two lines corresponding to the short and long trajectory groups. The red solid lines were calculated with an intermediate reciprocal intensity value ($\alpha_s^{\mathrm{H25}}<\alpha^{\mathrm{H25}}=1.67\times10^{-13}$~cm$^2$W$^{-1}<\alpha_l^{\mathrm{H25}}$) adjusted for best agreement with the fringe structure. The good overlap between these solid red lines (all nine calculated with the same $\alpha^{\mathrm{H25}}$ value) and the spectral interference pattern suggests that although the change in the instantaneous frequency of the fundamental beam causes apparent differences between the obtained textures in the vicinity of the 25\textsuperscript{th} harmonic, it leaves the $\alpha^q_j$ reciprocal intensities mainly unaffected. Closer to the cut-off regime of the spectrum, a single reciprocal intensity value ($\alpha^{\mathrm{H45}}=2.5\times10^{-13}$~cm$^2$W$^{-1}$, extracted from the Gábor-type transformation in figure \ref{Fig2}(c)) gives excellent agreement with the fringe pattern, as represented by the red dot-and-dashed lines in figures \ref{Fig3}(d) to (f)). During the investigation of similar interference features in spectrally and spatially resolved high-order harmonic radiation, Carlstr\"{o}m et al. also found that the curvature of the fringes depends on the mean value ($\alpha^q_l+\alpha^q_s)/2$, as well as on the difference $\Delta\alpha^q=\alpha^q_l-\alpha^q_s$, while the fringe periodicity is determined solely by $\Delta\alpha^q$. Utilizing this behaviour and a fitting procedure based on an analytical model, it was possible to retrieve the separate dipole phase parameters of the short ($\alpha^q_s$) and long ($\alpha^q_l$) trajectories from experimentally obtained modulation patterns \cite{Carlstrom2016}. It is important to note that the simple analytical model given in equation (\ref{Maker_fit}) does not take into account higher order terms in the intensity dependent dipole phase ($\mathcal{O}(\partial^2 \phi_j^q/\partial I^2)=0$), neither does it incorporate the different relative contributions of the trajectories. Such limitation is manifested in figure \ref{Fig3}(b), where the transition of the interference features from downward concave to downward convex at the harmonic centre is not represented well, despite the correct curvature prediction (with opposite orientation) by equation (\ref{Maker_fit}).

\section{The combined effect of the fundamental chirp}\label{Results_combined}
\begin{figure*}[t]
  \centering
  \includegraphics[width=0.95\linewidth]{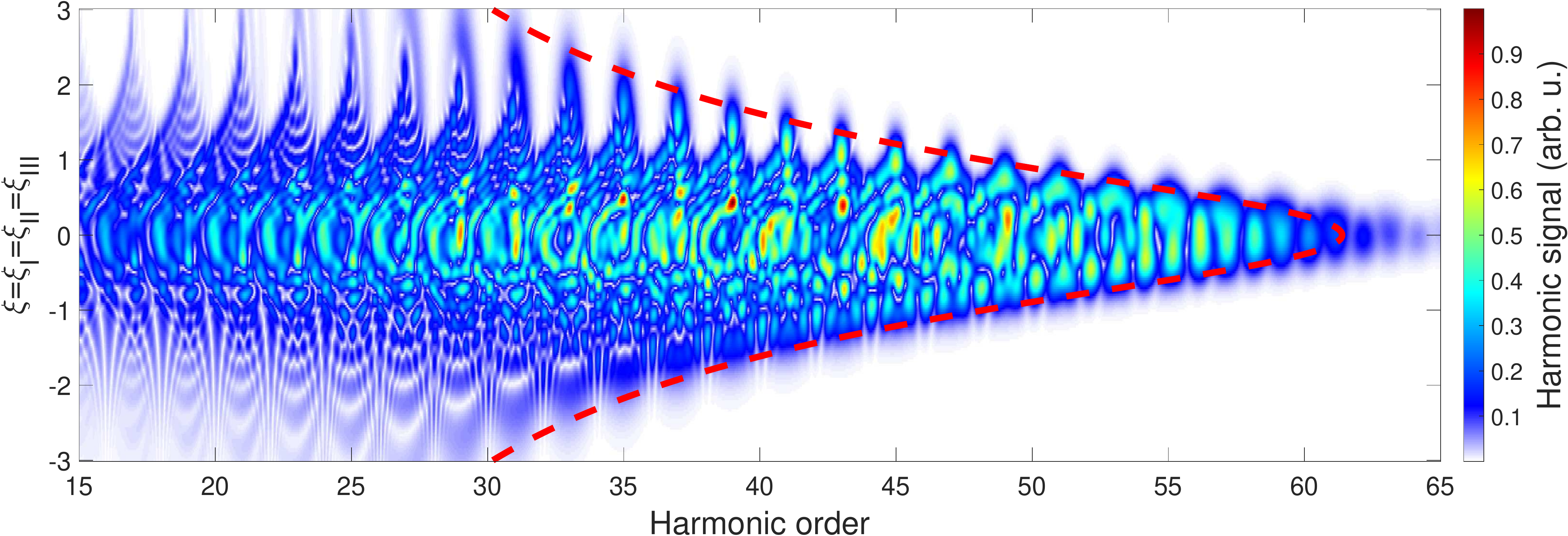}
  \caption{Progression of the single-atom harmonic spectra as a function of chirp parameter ($\xi=\xi_{\RN{1}}=\xi_{\RN{2}}=\xi_{\RN{3}}$). The dashed red line marks the calculated position of the cut-off. The bandwidth-limited pulse duration was $\tau_\mathrm{W,0}=15$~fs.}
  \label{Fig4}
\end{figure*}
In a real experimental scenario involving the change of the chirp (for example by rotating a plane-parallel plate, translating a wedge-pair or using an acousto-optical modulator), all three of the aforementioned processes must be simultaneously considered ($\xi=\xi_{\RN{1}}=\xi_{\RN{2}}=\xi_{\RN{3}}$ in (\ref{E_t_distinguished})). Figure~\ref{Fig4} reveals complicated interference structures on a spectrogram obtained by a chirp-scan: in case of short generating pulses (corresponding to chirp values from appr. $-0.5$ to $0.5$) the spectral map is dominated by erratic spectral fringes originating from the interferences between the small number of consecutive attosecond bursts. However, in the multi-cycle regime ($|\xi| > 0.5$), when a well-defined harmonic comb structure is formed, ripples analogous to the previously studied Maker fringes clearly emerge. In addition, features observable at even harmonic orders (in particular at negative $\xi$ values around $-2$) are similar to interferences between the long quantum paths only \cite{Sansone2006}. The combined effect of the harmonic chirp and the instantaneous frequency change of the fundamental field results in the firm sharpening of the harmonic streams for positively chirped generating pulses ($\xi > 0$), and wide, smoothed textures in case of negatively chirped drivers ($\xi < 0$). The cut-off position of each spectrum predicted by the Lewenstein model ($q_{\mathrm{cut-off}}=(F(I_\mathrm{p}/U_\mathrm{p})I_\mathrm{p}+3.17U_\mathrm{p})/\omega_0$, where $F(I_\mathrm{p}/U_\mathrm{p})$ is equal to 1.32 for $I_\mathrm{p}\ll U_\mathrm{p}$ and slowly reaches 1 as $I_\mathrm{p}/U_\mathrm{p}$ increases \cite{Lewenstein1994}) is marked with the dashed red line in figure \ref{Fig4}.\par
\begin{figure*}
  \centering
  \includegraphics[width=0.87\linewidth]{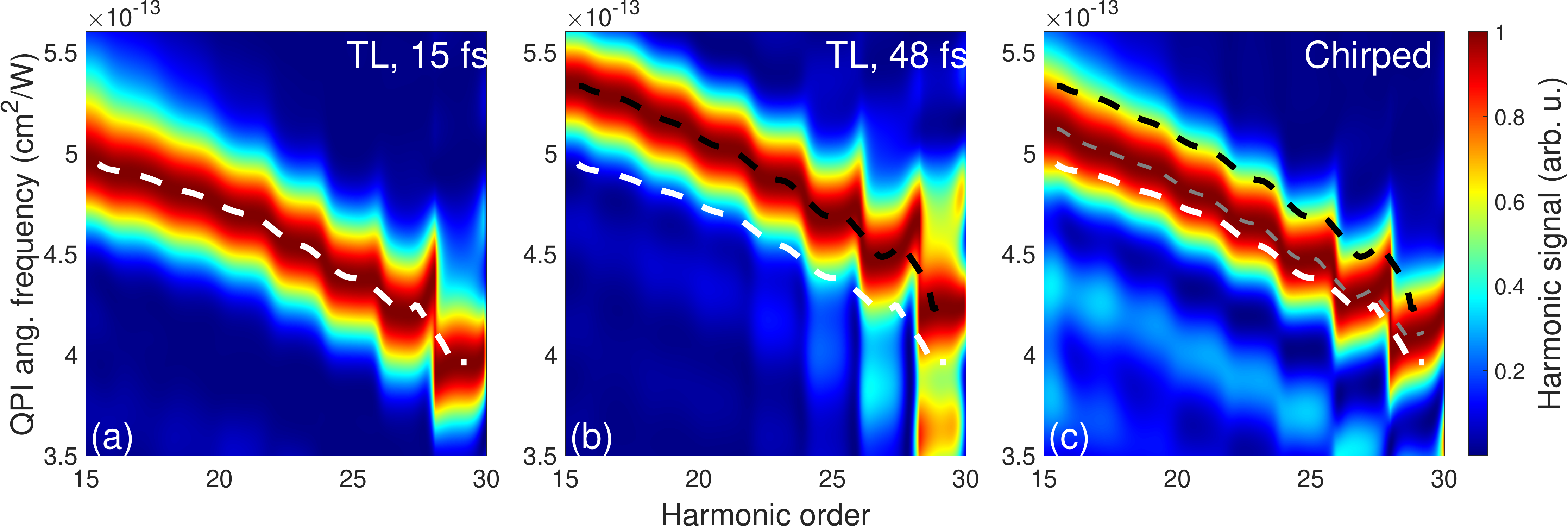}
  \caption{Density plots ($F(q,\alpha^q)$) obtained by the 2D Fourier analysis (equation (\ref{2D_Fourier})) of spectral maps produced with (a) transform-limited ($\xi_{{\RN{2}},{\RN{3}}}=0$, $|\xi_{\RN{1}}|\leq3$) 15~fs long, (b) transform-limited ($\xi_{\RN{2},\RN{3}}=0$, $|\xi_{\RN{1}}|\leq3$) 48~fs long and (c) chirped ($0\leq\xi=\xi_{\RN{1}}=\xi_{\RN{2}}=\xi_{\RN{3}}\leq3$) generating fields. The QPI frequencies at which the signal is maximum are marked by dashed white, black and grey lines for case (a), (b) and (c), respectively. $\lvert F(q,\alpha)\rvert^2$ is visualized, which was normalized independently for every $q$ harmonic order to enhance the visibility of the QPI signal.}
  \label{Fig5}
\end{figure*}

Although the QPI fringes are clearly visible on the spectral map for most plateau harmonics, evaluation of their periodicity is not straightforward due to the constantly varying pulse length, and the overlapping spectral interference patterns. Therefore, a 2D windowed Fourier transform was performed on the spectral maps ($S(q,\xi)$):
\begin{equation} \label{2D_Fourier}
F(q,\Delta\alpha)=\int\limits^{\infty}_{-\infty}\int\limits^{\infty}_{-\infty}e^{\frac{(q'-q)^2}{\Delta q^2}}S(q',\xi(I))dq'e^{-i\Delta\alpha I}dI ,
\end{equation}
where a window with the width of $\Delta q\approx$1 harmonic order was utilized (figure \ref{Fig5}). This technique was also used to estimate the error bars\textemdash for the points in figures \ref{Fig1}(b) and (c)\textemdash that were defined as the FWHM of the QPI angular frequency signal. The shift between QPI angular frequencies corresponding to different bandwidth-limited pulses\textemdash similarly to what was presented in figure \ref{Fig1}(b)\textemdash is clearly noticeable in figure \ref{Fig5}(a) and (b), when comparing the maxima of the angular frequency signals marked with white and black dashed lines for the transform-limited 15~fs and 48~fs driving pulses, respectively. Figure \ref{Fig5}(c) (obtained as the 2D windowed Fourier transform of figure \ref{Fig3}) shows a signal located between the white and black dashed lines corresponding to the 15 and 48~fs transform-limited cases, respectively. This is in agreement with the range of pulse durations in the chirped scenario spanning from 15~fs@$\xi=0$ to 48~fs@$\xi=3$ according to Table \ref{Parameters}.

\section{Macroscopic effects in QPI}
\label{macro}
\begin{figure*}
  \centering
  \includegraphics[width=1.0\linewidth]{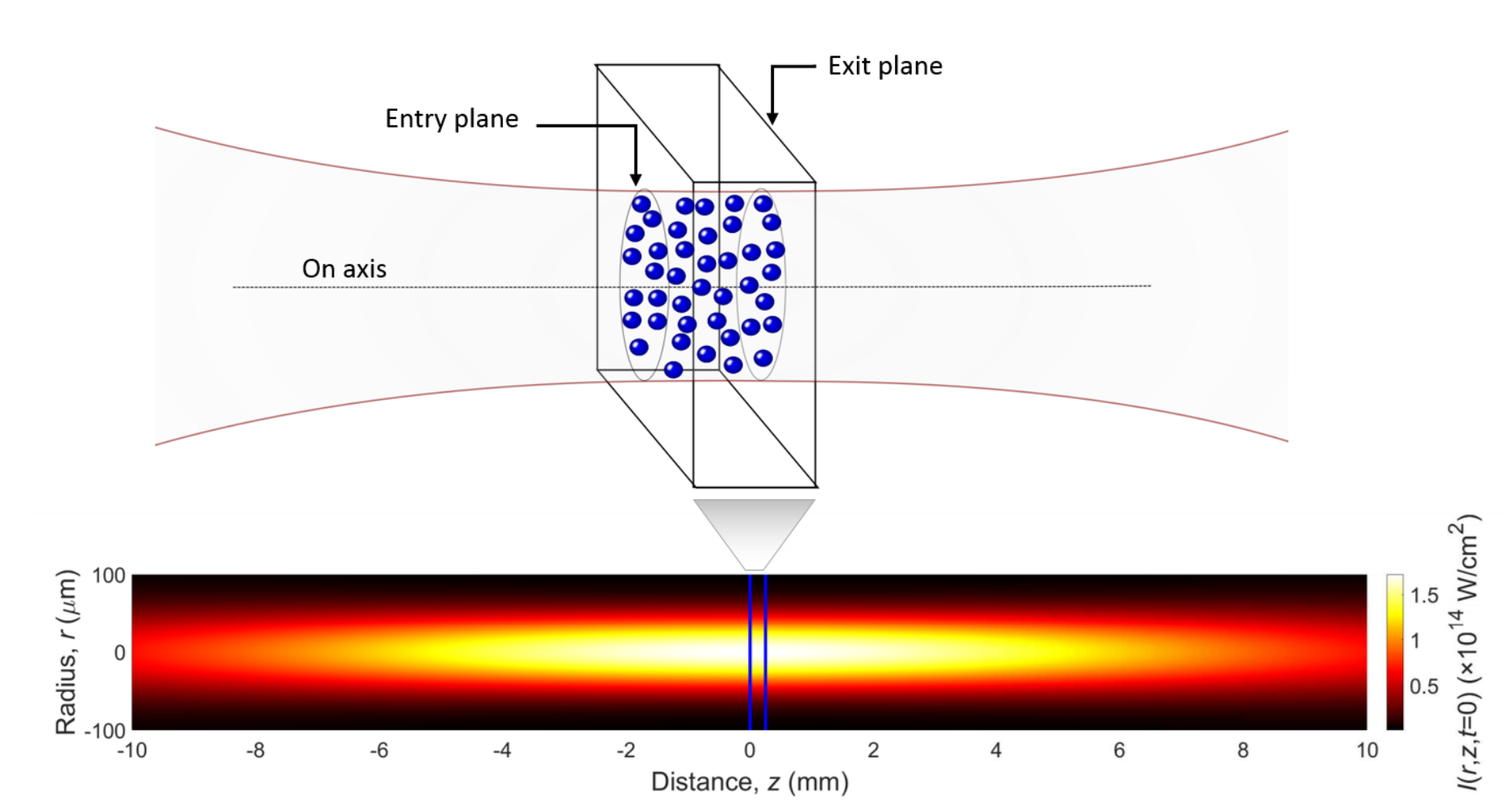}
  \caption{Schematic illustration of the focusing geometry used for macroscopic simulations. The entry and exit planes are demarcated. The colour map shows the peak intensity of the laser pulse as a function of the spatial coordinates. The blue lines mark the borders of the generation medium, showing that the confocal parameter is much larger than the jet size in this case.}
  \label{Fig6}
\end{figure*}

Finally, for a possible experimental observation of the discussed microscopic level interference textures, macroscopic effects (including plasma generation, absorption and refraction during propagation) cannot be neglected. We performed macroscopic simulations using a three-dimensional non-adiabatic model described in detail in Refs. \cite{Tosa2009,Priori2000,Major2019}.\par
The applied model consists of three main computational steps. In the first step, the propagation of the $E(\mathbf{r},t)$ electric field of the fundamental laser pulse ($E(\mathbf{r},t)=E(t)f(\mathbf{r})$, where $E(t)=-\partial{A(t)}/\partial{t}$) in the generating medium is calculated by solving the nonlinear wave equation \cite{Tosa2009}
\begin{equation}
\nabla^2E(\mathbf{r},t)-\frac{1}{c^2}\frac{\partial^2E(\mathbf{r},t)}{\partial t^2}=\frac{\omega_0^2}{c^2}(1-\eta_\mathrm{eff}^2(\mathbf{r},t))E(\mathbf{r},t) ,
\label{WaveEqn1}
\end{equation}
where $c$ is the speed of light in vacuum. The effective refractive index $\eta_\mathrm{eff}(\mathbf{r},t)$ is calculated as \cite{Tosa2016}
\begin{equation}
\eta_\mathrm{eff}(\mathbf{r},t)=\eta_0+\bar{\eta}_2 I(\mathbf{r},t)-\frac{\omega_p^2(\mathbf{r},t)}{2\omega_\mathrm{0}^2},
\end{equation}
where $I(\mathbf{r},t)=\frac{1}{2}\epsilon_0c\lvert\tilde{E}(\mathbf{r},t)\rvert^2$ is the intensity envelope of the laser field (calculated using the complex electric field $\tilde{E}(\mathbf{r},t)$ \cite{Chang2011}), and $\omega_p^2=n_ee^2/(m\epsilon_0)$ is the square of the plasma frequency (expressed using the number density of electrons $n_e$, the elementary charge $e$, the effective mass of the electron $m$, and the vacuum permittivity $\epsilon_0$). Dispersion, absorption, Kerr effect, and plasma dispersion along with absorption losses due to ionization \cite{Geissler1999} are taken into account via the  linear ($\eta_0$) and nonlinear ($\bar{\eta}_2$) part of the refractive index, respectively. The model assumes cylindrical symmetry ($\mathbf{r}\rightarrow r,z$) and uses paraxial approximation \cite{Tosa2016}. By applying a moving frame with the speed of light, and by eliminating the temporal derivative using Fourier transform, the equation to be solved explicitly is
\begin{equation}
\left(\frac{\partial^2}{\partial r^2}+\frac{1}{r}\frac{\partial}{\partial r}\right)E(r,z,\omega)-\frac{2i\omega}{c}\frac{\partial E(r,z,\omega)}{\partial z}=\frac{\omega^2}{c^2}\mathscr{F}[(1-\eta_\mathrm{eff}^2)E(\mathbf{r},t)] .
\end{equation}
The solution is obtained using the Crank–Nicolson method in an iterative algorithm \cite{Tosa2016}. The laser field distribution in the input plane of the medium is calculated using the ABCD-Hankel transformation \cite{Ibnchaikh2001}.\par
The second step is the calculation of the single-atom response (dipole moment $D(t)$) based on the laser-pulse temporal shapes available on the full ($r, z$) grid, using the Lewenstein integral defined in (\ref{Lew}). For the macroscopic nonlinear response $P_{nl}(t)$, the depletion of the ground state is taken into account \cite{Lewenstein1994,Le2009}, giving
\begin{equation}
    P_{nl}(t)=n_aD(t)\mathrm{exp}\bigg[-\int\limits^{t}_{-\infty}w(t')dt'\bigg] ,
\end{equation}
where $w(t)$ is the ionization rate obtained from tabulated values that were calculated using the hybrid anti-symmetrized coupled channels approach (haCC) \cite{Majety2015} showing good agreement with the Ammosov-Delone-Krainov (ADK) model \cite{ADK1986} and $n_a$ is the number density of atoms in the specific grid point ($r, z$) \cite{Tosa2016,Gaarde2008}.\par
The third and last step is to calculate the propagation of the generated radiation $E_h(\mathbf{r},t)$ by the wave equation
\begin{equation}
\nabla^2E_h(\mathbf{r},t)-\frac{1}{c^2}\frac{\partial^2E_h(\mathbf{r},t)}{\partial t^2}=\mu_0\frac{d^2P_{nl}(t)}{dt^2} ,
\label{WaveEqn2}
\end{equation}
with $\mu_0$ being the vacuum permeability. Equation (\ref{WaveEqn2}) is solved similarly to (\ref{WaveEqn1}), but since the source term is known, there is no need for an iterative scheme. The amplitude decrease and phase shift of the harmonic field - caused by absorption and dispersion, respectively - are taken into account at each step when solving equation (\ref{WaveEqn2}) using textbook expressions describing the effect of complex refractive index on wave propagation \cite{Hecht2017}. The real and imaginary parts of the refractive index in the XUV regime are calculated using tabulated values of atomic scattering factors \cite{Henke1993}.\par
\begin{figure*}
  \centering
  \includegraphics[width=1.0\linewidth]{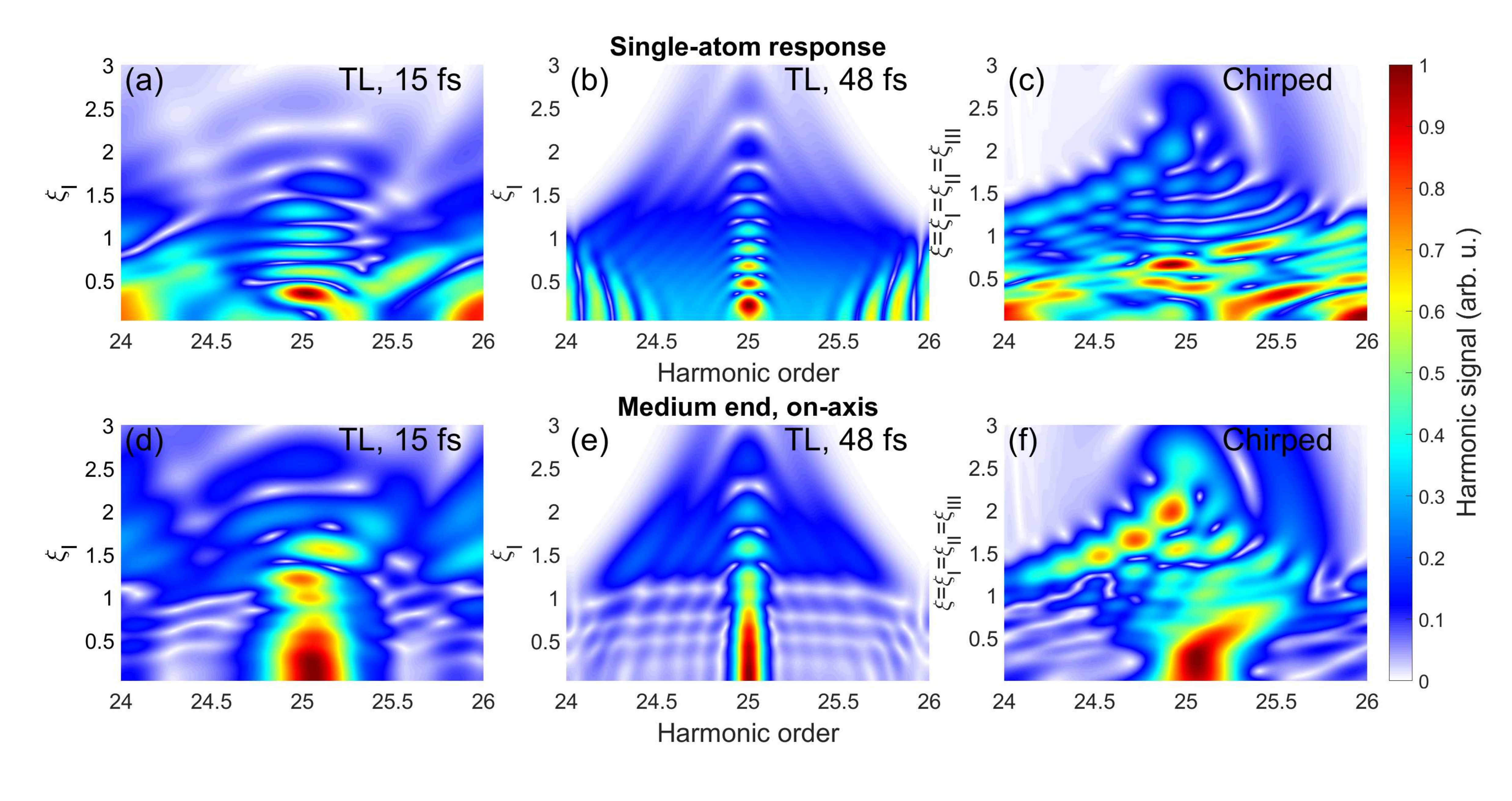}
  \caption{Evolution of the chirp-dependent harmonic spectra around the 25\textsuperscript{th} harmonic during macroscopic propagation in case of (a, d) transform-limited ($\xi_{{\RN{2}},{\RN{3}}}=0$, $|\xi_{\RN{1}}|\leq3$) 15~fs long, (b, e) transform-limited ($\xi_{\RN{2},\RN{3}}=0$, $|\xi_{\RN{1}}|\leq3$) 48~fs long and (c, f) chirped ($0\leq\xi=\xi_{\RN{1}}=\xi_{\RN{2}}=\xi_{\RN{3}}\leq3$) driving laser fields. (a--c) response at the beginning of the medium (equivalent to the single-atom response) corresponding to the entry plane of figure~\ref{Fig6}, (d--f) on-axis macroscopic result at the end of the medium corresponding to the exit plane of figure~\ref{Fig6}.}
  \label{Fig7}
\end{figure*}
It was assumed in the calculations that a 15~fs long pulse (in bandwidth-limit) having a Gaussian spatial intensity distribution with 5~mm intensity radius ($1/e^2$) and 187~$\mathrm{\mu J}$ total pulse energy is focused into argon gas with a 1~m focal length focusing optics (leading to a 66~$\mathrm{\mu m}$ focused beam waist) to match the parameter range used in the single-atom calculations (Table \ref{Parameters}) at the on-axis focal position. The beginning of the generating medium has been inserted at the position of the beam waist (figure~\ref{Fig6}). The gas medium had uniform pressure distribution of 667~Pa (5~Torr) and 250~$\mathrm{\mu m}$ length along the laser beam propagation axis.\par
Figure \ref{Fig7} compares the progression of spectral features around the 25\textsuperscript{th} harmonic as a function of the chirp parameter in case of the single-atom response (extracted at the beginning of the generation medium in the on-axis position, figure \ref{Fig7}(a--c)) and after macroscopic propagation (extracted at the end of the generation medium, also in the on-axis point, figure \ref{Fig7}(d--f)) using different, bandwidth-limited or chirped driving pulses. Such near field conditions can be experimentally monitored in multiple manners: for example the spatial XUV intensity distribution can be mapped onto a spatial ion distribution and recorded with an ion imaging detector (ion microscope) \cite{Kolliopoulos2014,Chatziathanasiou2019,Chatziathanasiou2019_2}. The atomic dipole phase can be directly measured by XUV-XUV interferometry \cite{Bellini1998, Corsi2006}, or the harmonic radiation can be detected by near field one-to-one imaging (see for example the experimental arrangements in refs. \cite{Ye2020, Hoflund2021}). It is shown in figure \ref{Fig7} that using the above specified, realistic set of macroscopic parameters results in similar microscopic and macroscopic patterns in case of 15 (a, d), and 48~fs (b, e) transform-limited pulses and for chirped (c, f) pulses as well, even though noticeable alterations naturally happen due to macroscopic propagation. This similarity can be attributed to the low level of ionization (see the previous footnote \ref{Footnote_ionization}), which induces only slight modifications of the temporal and spectral features of the composing pulses. The analogy between the microscopic and macroscopic cases is further supported by calculating the $\Delta\alpha^{\mathrm{H25}}$ angular frequencies of the QPI oscillations in all six instances (Table \ref{Macroscopic_vs_snglat_alphas}). Although the oscillation frequency is altered (in this case decreased) by the macroscopic propagation effects, the general tendency of the previous single-atom observations is maintained. The value of $\Delta\alpha^{\mathrm{H25}}$ increases with the increasing pulse duration of the transform-limited driving pulses from 15 to 48~fs, while a chirped driving pulse provides an intermediate $\Delta\alpha^{\mathrm{H25}}$ value.\par
The intrinsic intensity dependence of the phases of quantum paths establishes an analogy between spectral and spatial trajectory behaviour for analogous temporal and spatial field envelopes. Therefore, under careful experimental conditions, some of the findings in our paper might be reflected in the far field behaviour of trajectories. For example, the interference pattern in the divergence profiles might get modified, when the pulse duration is changed from the few to the multi-cycle regime due to the interplay between intra-pulse and spatial QPI patterns. To explore such effect, however, a more elaborated investigation is necessary that takes into account other important macroscopic contributions, such as interferences due to propagation, spatiotemporal coupling \cite{Dubrouil2014,Wikmark2019} or special effects entailing advanced pulse forms \cite{Raz2012}.
\begin{table}[b]
\centering
\begin{tabular}{|c|c|c|c|c|} 
\multicolumn{ 5}{c}{\textbf{QPI angular frequencies for the 25\textsuperscript{th} harmonic}} \\
\hline
\rowcolor[gray]{.9}[0.80\tabcolsep]
 & \textbf{TL, 15 fs} &  \textbf{TL, 48 fs} &  \textbf{chirped} & Unit\\ 
\hline\hline
\textbf{single atom}&4.47$\pm$0.14&4.74$\pm$0.12& 4.54$\pm$0.13&\multirow{2}{*}{$\times10^{-13}$ cm$^2$W$^{-1}$}\\
\cline{1-4}
\textbf{macroscopic, near field}&4.27$\pm$0.19&4.48$\pm$0.19&4.38$\pm$0.15&\\
\hline
\end{tabular}
\caption{Comparison of the $\Delta\alpha^{H25}$ values calculated from the modulation of the harmonic signal in case of the single atom, and near field macroscopic responses using figure \ref{Fig7} and equation (\ref{2D_Fourier}).}
\label{Macroscopic_vs_snglat_alphas}
\end{table}
\section{Discussion}\label{Discussion}
\begin{figure*}
  \centering
  \includegraphics[width=1.0\linewidth]{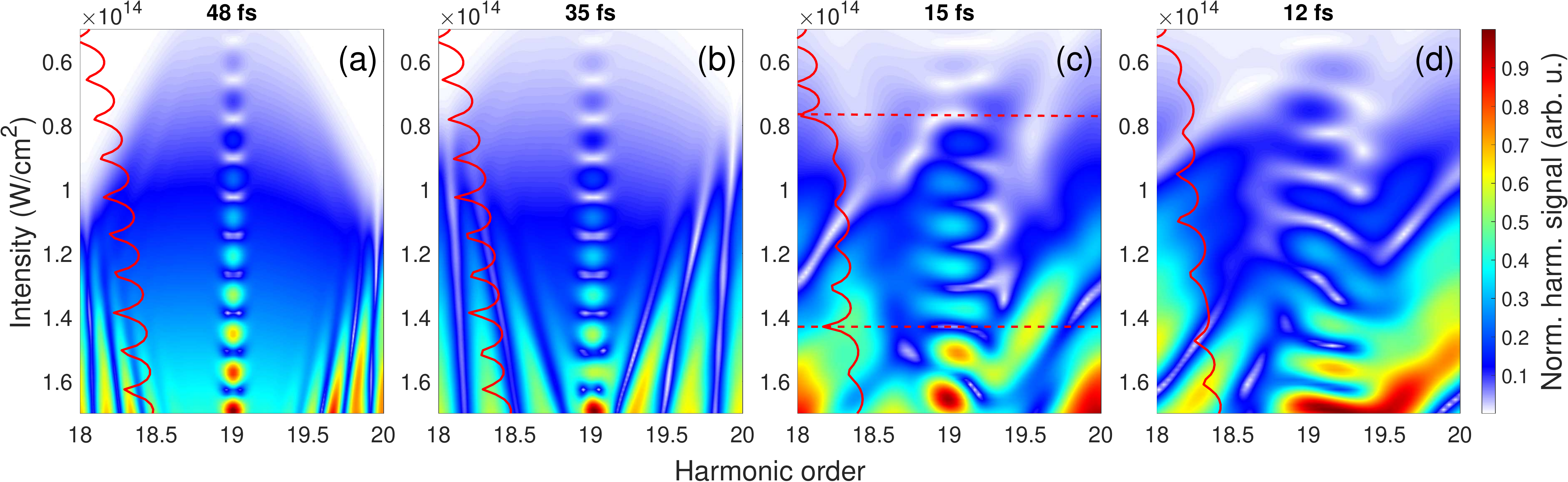}
  \caption{Detuning of the QPI oscillation frequency $\Delta\alpha^q$ of the 19$^{th}$ harmonic via the overlap between intra-pulse and intensity dependent QPI patterns. Fourier-limited driving pulse were utilized with the pulse duration $\tau_{\mathrm{W,0}}$ indicated in the title of each subplot. On the left side of each tile, the harmonic emission rate $S_q(\omega_q)$ evaluated at the spectral centre is plotted with a red solid line on a logarithmic scale. In (c), the position of two local dips in the QPI oscillation due to the overlap of interference minima was marked with red dashed lines.}
  \label{Fig8}
\end{figure*}

In our present study, quantum path interferences having various microscopic origins have been investigated. In the simplest case, one can consider the superposition of only two (one short, and one long) quantum trajectories within the same optical half-cycle of the laser field contributing to the same $q^\mathrm{th}$ harmonic of the fundamental laser frequency, causing \emph{intra-half-cycle} interference. For a real laser pulse containing multiple optical cycles, this plain picture is expanded by an interplay between trajectories in subsequent half-cycles (\emph{intra-pulse} interference) that forms spectral QPI patterns. Lastly, the phases of electron trajectories can be tuned pulse-to-pulse by varying the peak intensity of the fundamental laser field, leading to another interference-like pattern, which we refer to as \emph{intensity dependent} QPI pattern. In a macroscopic environment, this latter QPI pattern is mapped to spatially-resolved interferences stimulated by the spatial intensity distribution of the driving field profile \cite{Zair2008,Carlstrom2016,Heyl2011}. Although the complex, chirp (or intensity) dependent harmonic spectrum is formed by the inseparable superposition of these QPI patterns, our results indicate that by varying the driving pulse parameters, the domination of one or the other quantum level phase tuning mechanisms can be observed in the interference structure. This mechanism is illustrated in figure \ref{Fig8}. When, for example, the generating electric field is temporally long ($\tau_{W,0}>\approx$22~fs in our calculations, figure \ref{Fig8}(a) and (b)), the phase difference between long (short) paths in consecutive laser half-cycles is small, which induces constructive intra-pulse interference. This, in turn builds up narrow harmonics around the $q^\mathrm{th}$ harmonic order. In this case, the intensity-reciprocal intensity density map shows two, clearly separable lines (see figure~\ref{Fig2}(a)) corresponding to $\alpha^q_l$ and $\alpha^q_s$. This results in an intensity dependent QPI pattern with the observable frequency of $\Delta\alpha^q=\alpha^q_l-\alpha^q_s$ at harmonic order $q$. Contrarily, for very short driving pulses ($\tau_{W,0}<\approx$12~fs), the strong modulation of the phases of subsequent long (or short) trajectories through the half-cycle to half-cycle variation of the driving laser intensity creates destructive intra-pulse interference. This is manifested in an erratically behaving spectral QPI pattern, which washes out both the Gábor-type density plot, and the clear beating in the intensity dependent harmonic signal. A very interesting interference happens between these two pulse duration extremes ($\approx$12~fs$<\tau_{W,0}<\approx$22~fs, figure \ref{Fig8}(c) and (d)). While well-defined intensity dependent QPI pattern oscillation is still noticeable in this case, the intra-pulse modulation of the driving field detunes the phases of contributing trajectories in each laser-half-period, especially those of the long trajectories that (due to their longer $\tau^q_l$ excursion times) are more susceptible to the alterations of the electric field. This brings on the modification of the $\Delta\alpha^q$ oscillation frequency according to figure \ref{Fig1}(c). The superposition of spectral and intensity dependent QPI patterns is well marked in figure \ref{Fig7}(a) (calculated with transform-limited 15~fs fundamental pulses), and is also signed by the distortion of the interference pattern (by the occurance of two deeper minima at $\sim$0.8 and $1.5\times10^{14}$~W/cm$^2$ in the blue curve in figure \ref{Fig1}(a), also visualized by two horizontal dashed red lines in the corresponding figure \ref{Fig8}(c). On the other hand, in the 48~fs case (figure \ref{Fig7}(b)) with narrower spectral peaks, such interplay is shown only at the two sides of the 25$^{\mathrm{th}}$ harmonic (for $q<24.5$ and $q>25.5$), and no on-centre disturbance of peak intensity dependent oscillation can be observed.\par
In Section \ref{Results_TL} we found that the change of the instantaneous frequency within the driving pulse does not have a notable effect on the $\Delta\alpha^q$ angular frequency of the oscillation in the intensity dependent QPI pattern. Equation (\ref{Instfreq}) shows that the extent of instantaneous frequency change within a single laser burst cannot exceed the total optical bandwidth. Although the $\phi^q_j$ phase of a quantum path can be tuned by modifying the driving laser wavelength resulting in wavelength dependent QPI \cite{Schiessl2007}, this process would need a large spectral bandwidth, therefore only a few-cycle driving pulse to induce a notable effect. At the same time, destructive interferences for such short pulses would override clean intra-pulse oscillations.\par
Finally, simulations considering macroscopic propagation effects indicated that the interferences could be observed in carefully established experimental conditions yielding spectrograms similar to the ones presented in figure \ref{Fig7}. Such 3D spectral maps inherently contain information about pulse duration, which could be estimated by comparing the experimentally measured intensity dependent QPI oscillation periods to the calculated values. This \textit{in situ} information about the laser field on target during the extreme conditions of nonlinear interaction is difficult to obtain with conventional pulse characterization methods (like SPIDER \cite{Iaconis1998}, FROG \cite{Gallmann2000}, MIIPS \cite{Lozovoy2004}, etc.), either because of saturation and damage effects or practical constraints of the optical setup. The complete (pulse amplitude and phase) measurement \textit{in situ} at high laser power can be carried out using specific, recently developed techniques like THIS:d-scan \cite{Crespo2020} or in a more complex pump-probe scheme by attosecond streaking \cite{Goulielmakis2004}. The comparison of the measured pulse temporal properties and the duration estimates based on our discussion above may provide an experimentally simply obtained validation parameter for cross checking the correctness of such dedicated characterization techniques by utilizing the nonlinear nature of the HHG process itself.
\section{Conclusions}\label{Conclusions}
We have studied in detail the chirp dependence of quantum path interferences both on the single-atom level, as well as by macroscopic simulations, and demonstrated how a wide group of chirp-connected properties of the fundamental laser pulse, such as pulse duration, peak intensity and instantaneous frequency affect the quantum path interference patterns. We have shown that the periodicity of the interferometric beating introduced by the continuous change in the pulse peak intensity can be altered by the temporal extent of the driving field due to an interplay between different type of quantum path interferences. The presented calculations demonstrate the origin of a diversity of quantum path interference signatures that can be captured within the high harmonic spectra. Such analysis is essential for relating experimental measurements to microscopic interactions, and can be used as a guideline for precise control of ultrafast electron dynamics in gas phase HHG media.

\section{Acknowledgement}
The ELI-ALPS project (GINOP-2.3.6-15-2015-00001) is supported by the European Union and it is co-financed by the European Regional Development Fund. V.~T. acknowledges partial support from project  ELI\_03/01.10.2020 \textit{Pulse-MeReAd.} S.~K. acknowledges project No. 2018-2.1.14-T\'ET-CN-2018-00040, implemented with support provided by the National Research, Development and Innovation Fund of Hungary, and financed under the 2018-2.1.14-T\'ET-CN funding scheme. S.~K. also acknowledges project No. 2019-2.1.13-T\'ET-IN-2020-00059, which has been implemented with support provided by the National Research, Development and Innovation Fund of Hungary, and financed under the 2019-2.1.13-T\'ET-IN funding scheme. 
\section*{References}
\bibliographystyle{iopart-num}
\bibliography{References}

\providecommand{\newblock}{}
\begin{thebibliography}{10}
\expandafter\ifx\csname url\endcsname\relax
  \def\url#1{{\tt #1}}\fi
\expandafter\ifx\csname urlprefix\endcsname\relax\def\urlprefix{URL }\fi
\providecommand{\eprint}[2][]{\url{#2}}

\bibitem{Krausz2009}
Krausz F and Ivanov M 2009 {\em Rev. Mod. Phys.\/} {\bf 81} 163--264
  \urlprefix\url{https://doi.org/10.1103/RevModPhys.81.163}

\bibitem{Nayak2019}
Nayak A, Dumerque M, Kühn S, Mondal S, Csizmadia T, Harshitha N~G, Füle M,
  Kahaly M~U, Farkas B, Major B, Szaszkó-Bogár V, Földi P, Majorosi S,
  Tsatrafyllis N, Skantzakis E, Neoricic L, Shirozhan M, Vampa G, Varjú K,
  Tzallas P, Sansone G, Charalambidis D and Kahaly S 2019 {\em Phys. Rep.\/}
  {\bf 833} 1--52 \urlprefix\url{https://doi.org/10.1016/j.physrep.2019.10.002}

\bibitem{Maroju2020}
Maroju P~K, Grazioli C, Di~Fraia M, Moioli M, Ertel D, Ahmadi H, Plekan O,
  Finetti P, Allaria E, Giannessi L, De~Ninno G, Spezzani C, Penco G,
  Spampinati S, Demidovich A, Danailov M~B, Borghes R, Kourousias G, Sanches
  Dos~Reis C~E, Bill{\'e} F, Lutman A~A, Squibb R~J, Feifel R, Carpeggiani P,
  Reduzzi M, Mazza T, Meyer M, Bengtsson S, Ibrakovic N, Simpson E~R,
  Mauritsson J, Csizmadia T, Dumergue M, K{\"u}hn S, Nandiga~Gopalakrishna H,
  You D, Ueda K, Labeye M, B{\ae}kh{\o}j J~E, Schafer K~J, Gryzlova E~V,
  Grum-Grzhimailo A~N, Prince K~C, Callegari C and Sansone G 2020 {\em
  Nature\/} {\bf 578} 386--391
  \urlprefix\url{https://doi.org/10.1038/s41586-020-2005-6}

\bibitem{McPherson1987}
McPherson A, Gibson G, Jara H, Johann U, Luk T~S, McIntyre I~A, Boyer K and
  Rhodes C~K 1987 {\em J. Opt. Soc. Am. B\/} {\bf 4} 595--601
  \urlprefix\url{https://doi.org/10.1364/JOSAB.4.000595}

\bibitem{Ferray1988}
Ferray M, L'Huillier A, Li X~F, Lompre L~A, Mainfray G and Manus C 1988 {\em J.
  Phys. B: At. Mol. Opt. Phys.\/} {\bf 21} L31
  \urlprefix\url{https://doi.org/10.1088/0953-4075/21/3/001}

\bibitem{Kuhn2017}
Kühn S, Dumergue M, Kahaly S, Mondal S, Füle M, Csizmadia T, Farkas B, Major
  B, V{\'{a}}rallyay Z, Cormier E, Kalashnikov M, Calegari F, Devetta M,
  Frassetto F, M{\aa}nsson E, Poletto L, Stagira S, Vozzi C, Nisoli M, Rudawski
  P, Maclot S, Campi F, Wikmark H, Arnold C~L, Heyl C~M, Johnsson P, L'Huillier
  A, Lopez-Martens R, Haessler S, Bocoum M, Boehle F, Vernier A, Iaquaniello G,
  Skantzakis E, Papadakis N, Kalpouzos C, Tzallas P, L{\'{e}}pine F,
  Charalambidis D, Varj{\'{u}} K, Osvay K and Sansone G 2017 {\em J. Phys. B:
  At. Mol. Opt. Phys.\/} {\bf 50} 132002
  \urlprefix\url{https://doi.org/10.1088/1361-6455/aa6ee8}

\bibitem{Charalambidis2017}
Charalambidis D, Chik{\'{a}}n V, Cormier E, Dombi P, F{\"{u}}l{\"{o}}p J~A,
  Jan{\'{a}}ky C, Kahaly S, Kalashnikov M, Kamperidis C, K{\"{u}}hn S, Lepine
  F, L'Huillier A, Lopez-Martens R, Mondal S, Osvay K, {\'{O}}v{\'{a}}ri L,
  Rudawski P, Sansone G, Tzallas P, V{\'{a}}rallyay Z and Varj{\'{u}} K 2017
  {The Extreme Light Infrastructure—Attosecond Light Pulse Source (ELI-ALPS)
  Project} {\em Progress in Ultrafast Intense Laser Science XIII, Springer
  Series in Chemical Physics\/} ed Yamanouch K (Springer, Cham) pp 181--218
  \urlprefix\url{https://doi.org/10.1007/978-3-319-64840-8_10}

\bibitem{Major2018}
Major B, Farkas B, Dumergue M, Kovacs K, Kuhn S, L'Huillier A, Nagyill\'{e}s B,
  Rudawski P, Tosa V, Tzallas P, Charalambidis D, Osvay K, Sansone G and Varju
  K 2018 The eli alps research infrastructure: Scaling attosecond pulse
  generation for a large scale infrastructure {\em High-Brightness Sources and
  Light-driven Interactions\/} (Opt. Soc. Am.) p HW4A.1
  \urlprefix\url{https://doi.org/10.1364/HILAS.2018.HW4A.1}

\bibitem{Ye2020}
Ye P, Csizmadia T, Oldal L~G, Gopalakrishna H~N, F\"{u}le M, Filus Z,
  Nagyill{\'{e}}s B, Div{\'{e}}ki Z, Gr{\'{o}}sz T, Dumergue M,
  J{\'{o}}j{\'{a}}rt P, Seres I, Bengery Z, Zuba V, V{\'{a}}rallyay Z, Major B,
  Frassetto F, Devetta M, Lucarelli G~D, Lucchini M, Moio B, Stagira S, Vozzi
  C, Poletto L, Nisoli M, Charalambidis D, Kahaly S, Zaïr A and Varj{\'{u}} K
  2020 {\em Journal of Physics B: Atomic, Molecular and Optical Physics\/} {\bf
  53} 154004 \urlprefix\url{https://doi.org/10.1088/1361-6455/ab92bf}

\bibitem{Paul2001}
Paul P~M, Toma E~S, Breger P, Mullot G, Aug{\'e} F, Balcou P, Muller H~G and
  Agostini P 2001 {\em Science\/} {\bf 292} 1689--1692
  \urlprefix\url{https://doi.org/10.1126/science.1059413}

\bibitem{Carrera2006}
Carrera J~J, Tong X~M and Chu S~I 2006 {\em Phys. Rev. A\/} {\bf 74}(2) 023404
  \urlprefix\url{https://doi.org/10.1103/PhysRevA.74.023404}

\bibitem{Goulielmakis2008}
Goulielmakis E, Schultze M, Hofstetter M, Yakovlev V~S, Gagnon J, Uiberacker M,
  Aquila A~L, Gullikson E~M, Attwood D~T, Kienberger R, Krausz F and Kleineberg
  U 2008 {\em Science\/} {\bf 320} 1614--1617
  \urlprefix\url{https://doi.org/10.1126/science.1157846}

\bibitem{Lpine2013}
L{\'{e}}pine F, Sansone G and Vrakking M~J 2013 {\em Chemical Physics
  Letters\/} {\bf 578} 1--14
  \urlprefix\url{https://doi.org/10.1016/j.cplett.2013.05.045}

\bibitem{Kim2014}
Kim K~T, Villeneuve D~M and Corkum P~B 2014 {\em Nature Photonics\/} {\bf 8}
  187--194 \urlprefix\url{https://doi.org/10.1038/nphoton.2014.26}

\bibitem{Orfanos2020}
Orfanos I, Makos I, Liontos I, Skantzakis E, Major B, Nayak A, Dumergue M,
  K\"{u}hn S, Kahaly S, Varju K, Sansone G, Witzel B, Kalpouzos C, Nikolopoulos
  L~A~A, Tzallas P and Charalambidis D 2020 {\em Journal of Physics:
  Photonics\/} {\bf 2} 042003
  \urlprefix\url{https://doi.org/10.1088/2515-7647/aba172}

\bibitem{Chatziathanasiou2017}
Chatziathanasiou S, Kahaly S, Skantzakis E, Sansone G, Lopez-Martens R,
  Haessler S, Varju K, Tsakiris G, Charalambidis D and Tzallas P 2017 {\em
  Photonics\/} {\bf 4} 26
  \urlprefix\url{https://doi.org/10.3390/photonics4020026}

\bibitem{Zhang1998}
Chang Z, Rundquist A, Wang H, Christov I, Kapteyn H~C and Murnane M~M 1998 {\em
  Phys. Rev. A\/} {\bf 58}(1) R30--R33
  \urlprefix\url{https://doi.org/10.1103/PhysRevA.58.R30}

\bibitem{Lee2001}
Lee D~G, Kim J~H, Hong K~H and Nam C~H 2001 {\em Phys. Rev. Lett.\/} {\bf 87}
  243902 \urlprefix\url{https://doi.org/10.1103/PhysRevLett.87.243902}

\bibitem{Astiaso2016}
Lara-Astiaso M, Silva R~E~F, Gubaydullin A, Rivi\`ere P, Meier C and
  Mart\'{\i}n F 2016 {\em Phys. Rev. Lett.\/} {\bf 117}(9) 093003
  \urlprefix\url{https://doi.org/10.1103/PhysRevLett.117.093003}

\bibitem{Peng2018}
Peng D, Frolov M~V, Pi L~W and Starace A~F 2018 {\em Phys. Rev. A\/} {\bf
  97}(5) 053414 \urlprefix\url{https://doi.org/10.1103/PhysRevA.97.053414}

\bibitem{Corkum1993}
Corkum P~B 1993 {\em Phys. Rev. Lett.\/} {\bf 71}(13) 1994--1997
  \urlprefix\url{https://doi.org/10.1103/PhysRevLett.71.1994}

\bibitem{Krause1992}
Krause J~L, Schafer K~J and Kulander K~C 1992 {\em Phys. Rev. Lett.\/} {\bf
  68}(24) 3535--3538
  \urlprefix\url{https://doi.org/10.1103/PhysRevLett.68.3535}

\bibitem{Lewenstein1994}
Lewenstein M, Balcou P, Ivanov M, L'Huillier A and Corkum P~B 1994 {\em Phys.
  Rev. A\/} {\bf 49} 2117--2132
  \urlprefix\url{https://doi.org/10.1103/PhysRevA.49.2117}

\bibitem{Sansone2006}
Sansone G, Benedetti E, Caumes J~P, Stagira S, Vozzi C, De~Silvestri S and
  Nisoli M 2006 {\em Phys. Rev. A\/} {\bf 73}(5) 053408
  \urlprefix\url{https://link.aps.org/doi/10.1103/PhysRevA.73.053408}

\bibitem{Sansone2004}
Sansone G, Vozzi C, Stagira S and Nisoli M 2004 {\em Phys. Rev. A\/} {\bf
  70}(1) 013411 \urlprefix\url{https://doi.org/10.1103/PhysRevA.70.013411}

\bibitem{Varju2009}
Varjú K, Johnsson P, Mauritsson J, L’Huillier A and López-Martens R 2009
  {\em American Journal of Physics\/} {\bf 77} 389--395
  \urlprefix\url{https://doi.org/10.1119/1.3086028}

\bibitem{Yost2009}
Yost D~C, Schibli T~R, Ye J, Tate J~L, Hostetter J, Gaarde M~B and Schafer K~J
  2009 {\em Nature Physics\/} {\bf 5} 815--820
  \urlprefix\url{https://doi.org/10.1038/nphys1398}

\bibitem{Pedatzur2015}
Pedatzur O, Orenstein G, Serbinenko V, Soifer H, Bruner B~D, Uzan A~J, Brambila
  D~S, Harvey A~G, Torlina L, Morales F, Smirnova O and Dudovich N 2015 {\em
  Nature Physics\/} {\bf 11} 815--819
  \urlprefix\url{https://doi.org/10.1038/nphys3436}

\bibitem{Azoury2018}
Azoury D, Kneller O, Rozen S, Bruner B~D, Clergerie A, Mairesse Y, Fabre B,
  Pons B, Dudovich N and Kr\"{u}ger M 2018 {\em Nature Photonics\/} {\bf 13}
  54--59 \urlprefix\url{https://doi.org/10.1038/s41566-018-0303-4}

\bibitem{Zair2008}
Za\"{\i}r A, Holler M, Guandalini A, Schapper F, Biegert J, Gallmann L, Keller
  U, Wyatt A~S, Monmayrant A, Walmsley I~A, Cormier E, Auguste T, Caumes J~P
  and Sali\`eres P 2008 {\em Phys. Rev. Lett.\/} {\bf 100}(14) 143902
  \urlprefix\url{https://doi.org/10.1103/PhysRevLett.100.143902}

\bibitem{Schiessl2007}
Schiessl K, Ishikawa K~L, Persson E and Burgd\"orfer J 2007 {\em Phys. Rev.
  Lett.\/} {\bf 99}(25) 253903
  \urlprefix\url{https://doi.org/10.1103/PhysRevLett.99.253903}

\bibitem{Kanai2005}
Kanai T, Minemoto S and Sakai H 2005 {\em Nature\/} {\bf 435} 470--474
  \urlprefix\url{https://doi.org/10.1038/nature03577}

\bibitem{Cormier2009}
Cormier E, Za{\"i}r A, Holler M, Schapper F, Keller U, Wyatt A, Monmayrant A,
  Walmsley I, Auguste T and Sali{\`e}res P 2009 {\em Eur. Phys. J. ST\/} {\bf
  175} 191--194 \urlprefix\url{https://doi.org/10.1140/epjst/e2009-01140-5}

\bibitem{Seres2015}
Seres J, Seres E, Landgraf B, Aurand B, Kuehl T and Spielmann C 2015 {\em
  Photonics\/} {\bf 2} 104--123 ISSN 2304-6732
  \urlprefix\url{https://doi.org/10.3390/photonics2010104}

\bibitem{Kolliopoulos2014}
Kolliopoulos G, Bergues B, Schr\"oder H, Carpeggiani P~A, Veisz L, Tsakiris
  G~D, Charalambidis D and Tzallas P 2014 {\em Phys. Rev. A\/} {\bf 90}(1)
  013822 \urlprefix\url{https://doi.org/10.1103/PhysRevA.90.013822}

\bibitem{Chatziathanasiou2019}
Chatziathanasiou S, Kahaly S, Charalambidis D, Tzallas P and Skantzakis E 2019
  {\em Opt. Express\/} {\bf 27} 9733--9739
  \urlprefix\url{https://doi.org/10.1364/OE.27.009733}

\bibitem{Chatziathanasiou2019_2}
Chatziathanasiou S, Liontos I, Skantzakis E, Kahaly S, Kahaly M~U, Tsatrafyllis
  N, Faucher O, Witzel B, Papadakis N, Charalambidis D and Tzallas P 2019 {\em
  Phys. Rev. A\/} {\bf 100}(6) 061404
  \urlprefix\url{https://doi.org/10.1103/PhysRevA.100.061404}

\bibitem{Liu2009}
Liu C, Zheng Y, Zeng Z, Liu P, Li R and Xu Z 2009 {\em Opt. Express\/} {\bf 17}
  10319--10326 \urlprefix\url{https://doi.org/10.1364/OE.17.010319}

\bibitem{Carlstrom2016}
Carlström S, Precl{\'{\i}}kov{\'{a}} J, Lorek E, Larsen E~W, Heyl C~M,
  Pale{\v{c}}ek D, Zigmantas D, Schafer K~J, Gaarde M~B and Mauritsson J 2016
  {\em New J. Phys.\/} {\bf 18} 123032
  \urlprefix\url{https://doi.org/10.1088/1367-2630/aa511f}

\bibitem{Haessler2014}
Haessler S, Bal\ifmmode~\check{c}\else \v{c}\fi{}iunas T, Fan G, Andriukaitis
  G, Pug\ifmmode~\check{z}\else \v{z}\fi{}lys A, Baltu\ifmmode~\check{s}\else
  \v{s}\fi{}ka A, Witting T, Squibb R, Za\"{\i}r A, Tisch J~W~G, Marangos J~P
  and Chipperfield L~E 2014 {\em Phys. Rev. X\/} {\bf 4}(2) 021028
  \urlprefix\url{https://link.aps.org/doi/10.1103/PhysRevX.4.021028}

\bibitem{Praxmeyer2005}
Praxmeyer L and Wodkiewicz K 2005 {\em Laser Physics\/} {\bf 15} 1477
  (\textit{Preprint} \eprint{arXiv:physics/0502079})

\bibitem{Diels2006}
Diels J~C and Rudolph W 2006 Fundamentals, techniques, and applications on a
  femtosecond time scale {\em Ultrashort Laser Pulse Phenomena (Second
  Edition)\/} ed Diels J~C and Rudolph W (Burlington: Academic Press) pp 1 --
  60 second edition ed ISBN 978-0-12-215493-5
  \urlprefix\url{https://doi.org/10.1016/B978-012215493-5/50002-1}

\bibitem{Nakajima2007}
Nakajima T 2007 {\em Phys. Rev. A\/} {\bf 75}(5) 053409
  \urlprefix\url{https://doi.org/10.1103/PhysRevA.75.053409}

\bibitem{NakajimaCormier2007}
Nakajima T and Cormier E 2007 {\em Opt. Lett.\/} {\bf 32} 2879--2881
  \urlprefix\url{https://doi.org/10.1364/OL.32.002879}

\bibitem{Peng2009}
Peng L~Y, Tan F, Gong Q, Pronin E~A and Starace A~F 2009 {\em Phys. Rev. A\/}
  {\bf 80}(1) 013407 \urlprefix\url{https://doi.org/10.1103/PhysRevA.80.013407}

\bibitem{Mackenroth2016_PRL}
Mackenroth F, Gonoskov A and Marklund M 2016 {\em Phys. Rev. Lett.\/} {\bf
  117}(10) 104801
  \urlprefix\url{https://doi.org/10.1103/PhysRevLett.117.104801}

\bibitem{wahlstrom1993}
Wahlstr{\"{o}}m C~G, Larsson J, Persson A, Starczewski T, Svanberg S,
  Sali{\`{e}}res P, Balcou P and L'Huillier A 1993 {\em Physical Review A\/}
  {\bf 48} 4709--4720 \urlprefix\url{https://doi.org/10.1103/PhysRevA.48.4709}

\bibitem{Minemoto2008}
Minemoto S, Umegaki T, Oguchi Y, Morishita T, Le A~T, Watanabe S and Sakai H
  2008 {\em Physical Review A\/} {\bf 78} 061402
  \urlprefix\url{https://doi.org/10.1103/PhysRevA.78.061402}

\bibitem{Worner2009b}
W{\"{o}}rner H~J, Niikura H, Bertrand J~B, Corkum P~B and Villeneuve D~M 2009
  {\em Physical Review Letters\/} {\bf 102} 103901
  \urlprefix\url{https://doi.org/10.1103/PhysRevLett.102.103901}

\bibitem{Hassouneh2018}
Hassouneh O, Tyndall N~B, Wragg J, {Van Der Hart} H~W and Brown A~C 2018 {\em
  Physical Review A\/} {\bf 98} 1--9
  \urlprefix\url{https://doi.org/10.1103/PhysRevA.98.043419}

\bibitem{Scarborough2018}
Scarborough T, Gorman T, Mauger F, S{\'{a}}ndor P, Khatri S, Gaarde M, Schafer
  K, Agostini P and DiMauro L 2018 {\em Applied Sciences\/} {\bf 8} 1129
  \urlprefix\url{https://doi.org/10.3390/app8071129}

\bibitem{Colosimo2008}
Colosimo P, Doumy G, Blaga C~I, Wheeler J, Hauri C, Catoire F, Tate J, Chirla
  R, March A~M, Paulus G~G, Muller H~G, Agostini P and DiMauro L~F 2008 {\em
  Nature Physics\/} {\bf 4} 386--389
  \urlprefix\url{https://doi.org/10.1038/nphys914}

\bibitem{Schoun2014}
Schoun S~B, Chirla R, Wheeler J, Roedig C, Agostini P, DiMauro L~F, Schafer K~J
  and Gaarde M~B 2014 {\em Physical Review Letters\/} {\bf 112} 153001
  (\textit{Preprint} \eprint{1310.7008})
  \urlprefix\url{https://doi.org/10.1103/PhysRevLett.112.153001}

\bibitem{Shiner2011b}
Shiner A~D, Schmidt B~E, Trallero-Herrero C, W{\"{o}}rner H~J, Patchkovskii S,
  Corkum P~B, Kieffer J~C, L{\'{e}}gar{\'{e}} F and Villeneuve D~M 2011 {\em
  Nature Physics\/} {\bf 7} 464--467
  \urlprefix\url{https://doi.org/10.1038/nphys1940}

\bibitem{Zair2012}
Zaïr A, Siegel T, Sukiasyan S, Risoud F, Brugnera L, Hutchison C, Diveki Z,
  Auguste T, Tisch J~W, Salières P, Ivanov M~Y and Marangos J~P 2013 {\em
  Chemical Physics\/} {\bf 414} 184--191 ISSN 0301-0104 attosecond spectroscopy
  \urlprefix\url{https://doi.org/10.1016/j.chemphys.2012.12.022}

\bibitem{Schapper2010}
Schapper F, Holler M, Auguste T, Za\"{i}r A, Weger M, Sali\`{e}res P, Gallmann
  L and Keller U 2010 {\em Opt. Express\/} {\bf 18} 2987--2994
  \urlprefix\url{https://doi.org/10.1364/OE.18.002987}

\bibitem{Teichmann2012}
Teichmann S, Austin D, Bates P, Cousin S, Grün A, Clerici M, Lotti A, Faccio
  D, Trapani P~D, Couairon A and Biegert J 2012 {\em Laser Physics Letters\/}
  {\bf 9} 207--211 \urlprefix\url{https://doi.org/10.1002/lapl.201110116}

\bibitem{Majorosi2018}
Majorosi S, Benedict M~G and Czirj\'ak A 2018 {\em Phys. Rev. A\/} {\bf 98}(2)
  023401 \urlprefix\url{https://doi.org/10.1103/PhysRevA.98.023401}

\bibitem{He2015}
He L, Lan P, Zhang Q, Zhai C, Wang F, Shi W and Lu P 2015 {\em Phys. Rev. A\/}
  {\bf 92}(4) 043403 \urlprefix\url{https://doi.org/10.1103/PhysRevA.92.043403}

\bibitem{Varju2006}
Varj\'u K, Mairesse Y, Carr\'u B, Gaarde M~B, Johnsson P, Kazamias S,
  L\'opez-Martens R, Mauritsson J, Schafer K~J, Balcou P, L'Huillier A and
  Sali\`eres P 2006 {\em J. Mod. Opt.\/} {\bf 52} 379--394
  \urlprefix\url{https://doi.org/10.1080/09500340412331301542}

\bibitem{Balcou1999}
Balcou P, Dederichs A~S, Gaarde M~B and L’Huillier A 1999 {\em Journal of
  Physics B: Atomic, Molecular and Optical Physics\/} {\bf 32} 2973--2989
  \urlprefix\url{https://doi.org/10.1088/0953-4075/32/12/315}

\bibitem{Gaarde1999}
Gaarde M~B, Salin F, Constant E, Balcou P, Schafer K~J, Kulander K~C and
  L'Huillier A 1999 {\em Phys. Rev. A\/} {\bf 59}(2) 1367--1373
  \urlprefix\url{https://doi.org/10.1103/PhysRevA.59.1367}

\bibitem{Auguste2009}
Auguste T, Sali\`eres P, Wyatt A~S, Monmayrant A, Walmsley I~A, Cormier E,
  Za\"{\i}r A, Holler M, Guandalini A, Schapper F, Biegert J, Gallmann L and
  Keller U 2009 {\em Phys. Rev. A\/} {\bf 80}(3) 033817
  \urlprefix\url{https://doi.org/10.1103/PhysRevA.80.033817}

\bibitem{Nefedova2018}
Nefedova V~E, Ciappina M~F, Finke O, Albrecht M, V\'abek J, Kozlov\'a M,
  Su\'arez N, Pisanty E, Lewenstein M and Nejdl J 2018 {\em Phys. Rev. A\/}
  {\bf 98}(3) 033414 \urlprefix\url{https://doi.org/10.1103/PhysRevA.98.033414}

\bibitem{Salieres1995}
Sali\`eres P, L'Huillier A and Lewenstein M 1995 {\em Phys. Rev. Lett.\/} {\bf
  74}(19) 3776--3779
  \urlprefix\url{https://doi.org/10.1103/PhysRevLett.74.3776}

\bibitem{Kan1995}
Kan C, Capjack C~E, Rankin R and Burnett N~H 1995 {\em Phys. Rev. A\/} {\bf
  52}(6) R4336--R4339 \urlprefix\url{https://doi.org/10.1103/PhysRevA.52.R4336}

\bibitem{Kim2001}
Kim J~H, Lee D~G, Shin H~J and Nam C~H 2001 {\em Phys. Rev. A\/} {\bf 63}(6)
  063403 \urlprefix\url{https://doi.org/10.1103/PhysRevA.63.063403}

\bibitem{Shin1999}
Shin H~J, Lee D~G, Cha Y~H, Hong K~H and Nam C~H 1999 {\em Phys. Rev. Lett.\/}
  {\bf 83}(13) 2544--2547
  \urlprefix\url{https://doi.org/10.1103/PhysRevLett.83.2544}

\bibitem{Maker1962}
Maker P~D, Terhune R~W, Nisenoff M and Savage C~M 1962 {\em Phys. Rev. Lett.\/}
  {\bf 8}(1) 21--22 \urlprefix\url{https://doi.org/10.1103/PhysRevLett.8.21}

\bibitem{Heyl2011}
Heyl C~M, G\"udde J, H\"ofer U and L'Huillier A 2011 {\em Phys. Rev. Lett.\/}
  {\bf 107}(3) 033903
  \urlprefix\url{https://doi.org/10.1103/PhysRevLett.107.033903}

\bibitem{Catoire2016}
Catoire F, Ferr\'e A, Hort O, Dubrouil A, Quintard L, Descamps D, Petit S,
  Burgy F, M\'evel E, Mairesse Y and Constant E 2016 {\em Phys. Rev. A\/} {\bf
  94}(6) 063401 \urlprefix\url{https://doi.org/10.1103/PhysRevA.94.063401}

\bibitem{Tosa2009}
Tosa V, Kim K~T and Nam C~H 2009 {\em Phys. Rev. A\/} {\bf 79}(4) 043828
  \urlprefix\url{https://doi.org/10.1103/PhysRevA.79.043828}

\bibitem{Priori2000}
Priori E, Cerullo G, Nisoli M, Stagira S, De~Silvestri S, Villoresi P, Poletto
  L, Ceccherini P, Altucci C, Bruzzese R and de~Lisio C 2000 {\em Phys. Rev.
  A\/} {\bf 61}(6) 063801
  \urlprefix\url{https://doi.org/10.1103/PhysRevA.61.063801}

\bibitem{Major2019}
Major B, Kov\'{a}cs K, Tosa V, Rudawski P, L'Huillier A and Varj\'{u} K 2019
  {\em J. Opt. Soc. Am. B\/} {\bf 36} 1594--1601
  \urlprefix\url{https://doi.org/10.1364/JOSAB.36.001594}

\bibitem{Tosa2016}
Tosa V, Kov{\'{a}}cs K, Major B, Balogh E and Varj{\'{u}} K 2016 {\em Quantum
  Electronics\/} {\bf 46} 321--326
  \urlprefix\url{https://doi.org/10.1070/qel16039}

\bibitem{Chang2011}
Chang Z 2011 {\em Fundamentals of attosecond optics\/} (Boca Raton: CRC Press)
  p 547

\bibitem{Geissler1999}
Geissler M, Tempea G, Scrinzi A, Schn\"urer M, Krausz F and Brabec T 1999 {\em
  Phys. Rev. Lett.\/} {\bf 83}(15) 2930--2933
  \urlprefix\url{https://doi.org/10.1103/PhysRevLett.83.2930}

\bibitem{Ibnchaikh2001}
Ibnchaikh M and Belafhal A 2001 {\em Phys. Chem. News\/} {\bf 2} 29--34

\bibitem{Le2009}
Le A~T, Lucchese R~R, Tonzani S, Morishita T and Lin C~D 2009 {\em Phys. Rev.
  A\/} {\bf 80}(1) 013401
  \urlprefix\url{https://doi.org/10.1103/PhysRevA.80.013401}

\bibitem{Majety2015}
Majety V~P and Scrinzi A 2015 {\em Journal of Physics B: Atomic, Molecular and
  Optical Physics\/} {\bf 48} 245603
  \urlprefix\url{https://doi.org/10.1088/0953-4075/48/24/245603}

\bibitem{ADK1986}
Ammosov M~V, Delone N~B and Krainov V~P 1987 {\em Sov. Phys. JETP\/} {\bf 64}
  1191

\bibitem{Gaarde2008}
Gaarde M~B, Tate J~L and Schafer K~J 2008 {\em Journal of Physics B: Atomic,
  Molecular and Optical Physics\/} {\bf 41} 132001
  \urlprefix\url{https://doi.org/10.1088/0953-4075/41/13/132001}

\bibitem{Hecht2017}
Hecht E 2017 {\em Optics\/} (Boston: Addison-Wesley) p 728

\bibitem{Henke1993}
Henke B, Gullikson E and Davis J 1993 {\em Atomic Data and Nuclear Data
  Tables\/} {\bf 54} 181--342
  \urlprefix\url{https://doi.org/10.1006/adnd.1993.1013}

\bibitem{Bellini1998}
Bellini M, Lyng\aa{} C, Tozzi A, Gaarde M~B, H\"ansch T~W, L'Huillier A and
  Wahlstr\"om C~G 1998 {\em Phys. Rev. Lett.\/} {\bf 81}(2) 297--300
  \urlprefix\url{https://doi.org/10.1103/PhysRevLett.81.297}

\bibitem{Corsi2006}
Corsi C, Pirri A, Sali E, Tortora A and Bellini M 2006 {\em Phys. Rev. Lett.\/}
  {\bf 97}(2) 023901
  \urlprefix\url{https://doi.org/10.1103/PhysRevLett.97.023901}

\bibitem{Hoflund2021}
Hoflund M, Peschel J, Plach M, Dacasa H, Veyrinas K, Constant E, Smorenburg P,
  Wikmark H, Maclot S, Guo C, Arnold C, L'Huillier A and Eng-Johnsson P 2021
  {\em Ultrafast Science\/} {\bf 2021} 9797453
  \urlprefix\url{https://doi.org/10.34133/2021/9797453}

\bibitem{Dubrouil2014}
Dubrouil A, Hort O, Catoire F, Descamps D, Petit S, M{\'e}vel E, Strelkov V~V
  and Constant E 2014 {\em Nature Communications\/} {\bf 5} 4637
  \urlprefix\url{https://doi.org/10.1038/ncomms5637}

\bibitem{Wikmark2019}
Wikmark H, Guo C, Vogelsang J, Smorenburg P~W, Coudert-Alteirac H, Lahl J,
  Peschel J, Rudawski P, Dacasa H, Carlstr{\"o}m S, Maclot S, Gaarde M~B,
  Johnsson P, Arnold C~L and L{\textquoteright}Huillier A 2019 {\em Proceedings
  of the National Academy of Sciences\/} {\bf 116} 4779--4787
  \urlprefix\url{https://doi.org/10.1073/pnas.1817626116}

\bibitem{Raz2012}
Raz O, Pedatzur O, Bruner B~D and Dudovich N 2012 {\em Nature Photonics\/} {\bf
  6} 170--173 \urlprefix\url{https://doi.org/10.1038/nphoton.2011.353}

\bibitem{Iaconis1998}
Iaconis C and Walmsley I~A 1998 {\em Opt. Lett.\/} {\bf 23} 792--794
  \urlprefix\url{https://doi.org/10.1364/OL.23.000792}

\bibitem{Gallmann2000}
Gallmann L, Sutter D~H, Matuschek N, Steinmeyer G and Keller U 2000 {\em
  Applied Physics B\/} {\bf 70} S67--S75
  \urlprefix\url{https://doi.org/10.1007/s003400000307}

\bibitem{Lozovoy2004}
Lozovoy V~V, Pastirk I and Dantus M 2004 {\em Opt. Lett.\/} {\bf 29} 775--777
  \urlprefix\url{https://doi.org/10.1364/OL.29.000775}

\bibitem{Crespo2020}
Crespo H~M, Witting T, Canhota M, Miranda M and Tisch J~W~G 2020 {\em Optica\/}
  {\bf 7} 995--1002 \urlprefix\url{https://doi.org/10.1364/OPTICA.398319}

\bibitem{Goulielmakis2004}
Goulielmakis E, Uiberacker M, Kienberger R, Baltuska A, Yakovlev V, Scrinzi A,
  Westerwalbesloh T, Kleineberg U, Heinzmann U, Drescher M and Krausz F 2004
  {\em Science\/} {\bf 305} 1267--1269 ISSN 0036-8075
  \urlprefix\url{https://doi.org/10.1126/science.1100866}

\end{thebibliography}
\end{document}